\documentclass[english,aps,superscriptaddress]{revtex4}
\usepackage[T1]{fontenc}
\usepackage[latin9]{inputenc}
\usepackage{mathrsfs}
\usepackage{amsbsy}
\usepackage{esint}

\makeatletter
\@ifundefined{textcolor}{}
{%
 \definecolor{BLACK}{gray}{0}
 \definecolor{WHITE}{gray}{1}
 \definecolor{RED}{rgb}{1,0,0}
 \definecolor{GREEN}{rgb}{0,1,0}
 \definecolor{BLUE}{rgb}{0,0,1}
 \definecolor{CYAN}{cmyk}{1,0,0,0}
 \definecolor{MAGENTA}{cmyk}{0,1,0,0}
 \definecolor{YELLOW}{cmyk}{0,0,1,0}
}

\makeatother

\usepackage{babel}
\begin{document}

\title{Covariant chiral kinetic equation in Wigner function approach}

\author{Jian-hua Gao}

\affiliation{Shandong Provincial Key Laboratory of Optical Astronomy and Solar-Terrestrial
Environment, Institute of Space Sciences, Shandong University, Weihai,
Shandong 264209, China}

\author{Shi Pu}

\affiliation{Department of Physics, The University of Tokyo, 7-3-1 Hongo, Bunkyo-ku,
Tokyo 113-0033, Japan}

\author{Qun Wang}

\affiliation{Department of Modern Physics, University of Science and Technology
of China, Hefei, Anhui 230026, China}
\begin{abstract}
The covariant chiral kinetic equation (CCKE) is derived from the 4-dimensional
Wigner function by an improved perturbative method under the static
equilibrium conditions. The chiral kinetic equation in 3-dimensions
can be obtained by integration over the time component of the 4-momentum.
There is freedom to add more terms to the CCKE allowed by conservation
laws. In the derivation of the 3-dimensional equation, there is also
freedom to choose coefficients of some terms in $dx_{0}/d\tau$ and
$d\mathbf{x}/d\tau$ ($\tau$ is a parameter along the worldline,
and $(x_{0},\mathbf{x})$ denotes the time-space position of a particle)
whose 3-momentum integrals are vanishing. So the 3-dimensional chiral
kinetic equation derived from the CCKE is not uniquely determined
in the current approach. To go beyond the current approach, one needs
a new way of building up the 3-dimensional chiral kinetic equation
from the CCKE or directly from covariant Wigner equations. 
\end{abstract}
\maketitle

\section{Introduction}

The chiral or axial vector anomaly is the anomalous nonconservation
of a chiral or axial vector current of fermions arising from quantum
effects, it is also called Adler-Bell-Jackiw (ABJ) anomaly \cite{Adler:1969gk,Bell:1969ts}.
The chiral anomaly manifests itself in the pion's decay into two photons,
which involves an AVV (axial-vector-vector) coupling with the chiral
current and the photon field being the axial vector and vector field
respectively. In magnetic fields, the chiral anomaly will lead to
an electric current along the magnetic field resulting from an imbalance
of chirality, which is called the chiral magnetic effect (CME) \cite{Vilenkin:1980fu,Kharzeev:2007jp,Fukushima:2008xe,Kharzeev:2015znc},
for reviews, see, e.g., Ref. \cite{Kharzeev:2012ph,Kharzeev:2013jha,Kharzeev:2015znc,Huang:2015oca}.
This effect gives another example of AVV coupling among the vector
current, the chiral chemical potential of fermions and the magnetic
field. The CME is also associated with the chiral vortical effect
in which an electric current is induced by the vorticity in a system
of charged particles \cite{Vilenkin:1978hb,Erdmenger:2008rm,Banerjee:2008th}.
In anomalous hydrodynamics the CME and CVE must coexist in order to
guarantee the second law of thermodynamics \cite{Son:2009tf,Pu:2010as,Ozonder:2010zy}.
The CME has recently been confirmed in materials such as Dirac and
Weyl semi-metals \cite{Son:2012bg,Basar:2013iaa,Li:2014bha}. 

In non-central heavy-ion collisions at high energies very strong magnetic
fields \cite{Kharzeev:2007jp,Skokov:2009qp,Voronyuk:2011jd,Deng:2012pc,Bloczynski:2012en,McLerran:2013hla,Gursoy:2014aka,Roy:2015coa,Tuchin:2014iua,Li:2016tel}
and huge global angular momenta \cite{Liang:2004ph,Liang:2004xn,Becattini:2007sr,Betz:2007kg,Gao:2007bc}
are produced. The CME, CVE and some other effects such as chiral magnetic
wave \cite{Kharzeev:2010gd,Burnier:2011bf} have been extensively
studied in heavy-ion collisions. The charge separation effect observed
in STAR \cite{Abelev:2009ac,Abelev:2009ad} and ALICE \cite{Abelev:2012pa}
experiments are consistent to the CME prediction. But there were debates
that the charge separation might arise from other effects such as
cluster particle correlations \cite{Wang:2009kd} or local charge
conservation \cite{Schlichting:2010qia}, so a substantial portion
of the two charged particle correlation measured in experiments may
come from backgrounds. Recently the CMS collaboration has measured
the the two charged particle correlation in pPb collisions \cite{Khachatryan:2016got}
and found a similar result to that of STAR \cite{Abelev:2009ac,Abelev:2009ad}
and ALICE \cite{Abelev:2012pa} in AuAu and PbPb collisions. The CMS
result supports that the measured azimuthal correlation of two charged
particles at TeV energies may probably come from backgrounds. More
efforts in theoretical and experimental investigation are needed to
separate the signal and backgrounds \cite{Sorensen:2017,Hirono:2014oda,Deng:2016knn}. 

Recently the STAR collaboration has measured a nonvanishing polarization
of $\Lambda$ and $\bar{\Lambda}$ hyperons along the global angular
momentum with respect of the reaction plane in its beam energy scan
program \cite{STAR:2017ckg}. This is a piece of evidence for the
local polarization effect from vorticity in collisions at lower energy
and was first predicted in Ref. \cite{Liang:2004ph} and later extensively
studied \cite{Becattini:2007sr,Betz:2007kg,Gao:2007bc,Becattini:2013fla,Pang:2016igs,Becattini:2016gvu,Xie:2016fjj,Karpenko:2016jyx,Aristova:2016wxe}. 

To describe the kinematics of chiral fermions with specific helicities
or massive fermions with spins, one needs to know their momenta as
well as their coordinates. Therefore people use quantum kinetic theory
in terms of Wigner function \cite{Heinz:1983nx,Elze:1986qd,Vasak:1987um,Zhuang:1995pd},
which turns out to be a useful tool to describe the CME, CVE, and
other related effects \cite{Gao:2012ix,Chen:2012ca,Gao:2015zka,Hidaka:2016yjf}.
The axial vector component of the Wigner function for massless fermions
can be generalized to massive fermions and gives their phase-space
density of the spin vector \cite{Fang:2016vpj}. The spin vector arises
from nonzero fermion mass \cite{Chen:2013iga}. Therefore one can
calculate the polarization of massive fermions from the axial vector
component, and the polarization density is found to be proportional
to the local vorticity $\boldsymbol{\omega}$ as well as the magnetic
field \cite{Fang:2016vpj}. 

An effective way of describing the kinematics of chiral fermions in
phase space in the presence of chiral anomaly is the chiral kinetic
equation \cite{Son:2012wh,Stephanov:2012ki,Chen:2012ca,Son:2012zy,Chen:2014cla,Yamamoto:2015gzz,Kharzeev:2016sut,Mueller:2017arw,Mueller:2017lzw,Hidaka:2016yjf},
which is closely related to the Berry phase and monopole in momentum
space. The covariant chiral kinetic equation (CCKE) can be derived
in the 4-dimensional (4D) Wigner function approach \cite{Chen:2012ca},
from which one can derive the 3-dimensional (3D) version of the chiral
kinetic equation by integration over the time component of the 4-momentum.
In this paper we will give a systematic derivation of the CCKE in
an improved method in comparison with Ref. \cite{Chen:2012ca}. We
will show how the 3D chiral kinetic equation can be derived from the
CCKE in details. The 3D chiral kinetic equation cannot be uniquely
determined in our current approach due to some free coefficients. 

The paper is organized as follows. In Section \ref{sec:wig} we introduce
the Wigner function and its solutions in static-equilibrium conditions.
In Section \ref{sec:ccke}, we give a systematic and improved derivation
of the CCKE in 4D. In Section \ref{sec:free}, we show that there
is freedom of adding more terms to the CCKE allowed by conservation
laws. In Section \ref{sec:4d-to-3d}, we derive the chiral kinetic
equation in 3D from the CCKE. In the final section, we give a summary
of the main results. 

We use the same sign convention for $Q$ and $\gamma_{5}$ as in Ref.
\cite{Gao:2012ix,Chen:2012ca}. The energy of a massless fermion with
a three-momentum $\mathbf{p}$ is denoted as $E_{p}=|\mathbf{p}|$.

\section{Wigner function and its solutions in static-equilibrium conditions}

\label{sec:wig}In a background electromagnetic field, the quantum
mechanical analogue of a classical phase-space distribution for fermions
is the gauge invariant Wigner function $W_{\alpha\beta}(x,p)$ which
satisfies the equation of motion \cite{Elze:1986qd,Vasak:1987um},
\begin{equation}
\left(\gamma_{\mu}K^{\mu}-m\right)W(x,p)=0
\end{equation}
where $x=(x_{0},\mathbf{x})$ and $p=(p_{0},\mathbf{p})$ are space-time
and energy-momentum 4-vectors. For the constant field strength $F_{\mu\nu}$,
the operator $K^{\mu}$ is given by $K^{\mu}=p^{\mu}+i\frac{1}{2}\nabla^{\mu}$
with $\nabla^{\mu}=\partial_{x}^{\mu}-QF^{\mu\nu}\partial_{\nu}^{p}$.
The Wigner function can be decomposed in 16 independent generators
of Clifford algebra, 
\begin{equation}
W=\frac{1}{4}\left[\mathscr{F}+i\gamma^{5}\mathscr{P}+\gamma^{\mu}\mathscr{V}_{\mu}+\gamma^{5}\gamma^{\mu}\mathscr{A}_{\mu}+\frac{1}{2}\sigma^{\mu\nu}\mathscr{S}_{\mu\nu}\right],\label{eq:wigner-decomp}
\end{equation}
whose coefficients $\mathscr{F}$, $\mathscr{P}$, $\mathscr{V}_{\mu}$,
$\mathscr{A}_{\mu}$ and $\mathscr{S}_{\mu\nu}$ are the scalar, pseudo-scalar,
vector, axial-vector and tensor components of the Wigner function
respectively. 

For massless or chiral fermions, the equations for $\mathscr{V}_{\mu}$
and $\mathscr{A}_{\mu}$ are decoupled from other components of the
Wigner function, from which one can obtain independent equations for
vector components $\mathscr{J}_{\mu}^{s}(x,p)$ of the Wigner function
for right-handed $(s=+)$ and left-handed ($s=-$) fermions, 
\begin{eqnarray}
p^{\mu}\mathscr{J}_{\mu}^{s}(x,p) & = & 0,\nonumber \\
\nabla^{\mu}\mathscr{J}_{\mu}^{s}(x,p) & = & 0,\nonumber \\
2s(p^{\lambda}\mathscr{J}_{s}^{\rho}-p^{\rho}\mathscr{J}_{s}^{\lambda}) & = & -\epsilon^{\mu\nu\lambda\rho}\nabla_{\mu}\mathscr{J}_{\nu}^{s},\label{eq:wig-eq}
\end{eqnarray}
where $\mathscr{J}_{\mu}^{s}(x,p)$ are defined as 
\begin{eqnarray}
\mathscr{J}_{\mu}^{s}(x,p) & = & \frac{1}{2}[\mathscr{V}_{\mu}(x,p)+s\mathscr{A}_{\mu}(x,p)].
\end{eqnarray}
One can derive a formal solution of $\mathscr{J}_{\mu}^{s}$ satisfying
Eq. (\ref{eq:wig-eq}) by a perturbation in powers of space-time derivative
$\partial_{\mu}^{x}$ and field strength $F_{\mu\nu}$. The solution
at the zeroth and first order reads 
\begin{eqnarray}
\mathscr{J}_{(0)s}^{\rho}(x,p) & = & p^{\rho}f_{s}\delta(p^{2}),\nonumber \\
\mathscr{J}_{(1)s}^{\rho}(x,p) & = & -\frac{s}{2}\tilde{\Omega}^{\rho\beta}p_{\beta}\frac{df_{s}}{dp_{0}}\delta(p^{2})-\frac{s}{p^{2}}Q\tilde{F}^{\rho\lambda}p_{\lambda}f_{s}\delta(p^{2}).\label{eq:1st-solution}
\end{eqnarray}
Here $p_{0}\equiv u\cdot p$ is the particle energy in the co-moving
frame of the fluid, $u^{\mu}$ is the fluid velocity, $\tilde{F}^{\rho\lambda}=\frac{1}{2}\epsilon^{\rho\lambda\mu\nu}F_{\mu\nu}$,
$\tilde{\Omega}^{\xi\eta}=\frac{1}{2}\epsilon^{\xi\eta\nu\sigma}\Omega_{\nu\sigma}$
with $\Omega_{\nu\sigma}=\frac{1}{2}(\partial_{\nu}u_{\sigma}-\partial_{\sigma}u_{\nu})$,
where $\epsilon^{\mu\nu\sigma\beta}$ and $\epsilon_{\mu\nu\sigma\beta}$
are anti-symmetric tensors with $\epsilon^{\mu\nu\sigma\beta}=1(-1)$
and $\epsilon_{\mu\nu\sigma\beta}=-1(1)$ for even (odd) permutations
of indices 0123, so we have $\epsilon^{0123}=-\epsilon_{0123}=1$.
Instead of $\Omega_{\nu\sigma}$, $\tilde{\Omega}^{\xi\eta}$, $F_{\mu\nu}$
and $\tilde{F}^{\rho\lambda}$, we will also use the vorticity vector
$\omega^{\rho}=\frac{1}{2}\epsilon^{\rho\sigma\alpha\beta}u_{\sigma}\partial_{\alpha}u_{\beta}$,
the electric field $E^{\mu}=F^{\mu\nu}u_{\nu}$, and the magnetic
field $B^{\mu}=\frac{1}{2}\epsilon^{\mu\nu\lambda\rho}u_{\nu}F_{\lambda\rho}$.
In Eq. (\ref{eq:1st-solution}) $f_{s}$ is the distribution function
for chiral fermions at the zeroth order, 
\begin{eqnarray}
f_{s}(x,p) & = & \frac{2}{(2\pi)^{3}}\left[\Theta(p_{0})f_{\mathrm{FD}}(p_{0}-\mu_{s})+\Theta(-p_{0})f_{\mathrm{FD}}(-p_{0}+\mu_{s})\right].\label{eq:dist}
\end{eqnarray}
Here $f_{\mathrm{FD}}(y)\equiv1/[\exp(\beta y)+1]$ is the Fermi-Dirac
distribution function, with $\beta=1/T$ and $\mu_{s}$ being the
temperature inverse and the chemical potential for chiral fermions
with chirality $s=\pm1$ respectively. We can express $\mu_{s}$ in
terms of the scalar and pseudo-scalar (or chiral) chemical potentials,
$\mu_{s}=\mu+s\mu_{5}$. 

The solution to the Wigner function in Eq. (\ref{eq:1st-solution})
is the result of following static-equilibrium conditions \cite{Gao:2012ix},
\begin{eqnarray}
 &  & \Delta^{\sigma\alpha}\Delta^{\rho\beta}\left(\partial_{\alpha}u_{\beta}+\partial_{\beta}u_{\alpha}-\frac{2}{3}\Delta_{\alpha\beta}\Delta^{\rho\sigma}\partial_{\rho}u_{\sigma}\right)=0,\nonumber \\
 &  & T\Delta^{\sigma\rho}\partial_{\rho}\frac{\mu}{T}+QE^{\sigma}=0,\nonumber \\
 &  & u^{\rho}\partial_{\rho}u^{\sigma}-\Delta^{\sigma\rho}\partial_{\rho}\ln T=0,\nonumber \\
 &  & \partial_{\sigma}\frac{\mu_{5}}{T}=0,\;\;\;\; u^{\sigma}\partial_{\sigma}\frac{\mu}{T}=0,\nonumber \\
 &  & u^{\sigma}\partial_{\sigma}T+\frac{1}{3}T\Delta^{\rho\sigma}\partial_{\rho}u_{\sigma}=0.
\end{eqnarray}
For simplicity we assume in this paper the equilibrium conditions
with constant temperature, then the above constraints are reduced
to the following conditions \cite{Gao:2012ix} 
\begin{eqnarray}
 &  & \partial_{\sigma}\mu_{5}=0,\quad\partial_{\sigma}T=0,\nonumber \\
 &  & \partial^{\rho}u^{\sigma}+\partial^{\sigma}u^{\rho}=0,\nonumber \\
 &  & \partial_{\sigma}\mu=-QE_{\sigma}.\label{eq:static-equil}
\end{eqnarray}
From above conditions, we can derive following identities $u_{\rho}\partial^{\rho}u^{\sigma}=0$,
$\partial_{\sigma}u^{\sigma}=0$, $\partial_{\mu}\omega^{\mu}=0$,
$\epsilon^{\mu\nu\alpha\beta}\omega_{\nu}u_{\alpha}B_{\beta}=0$,
$\Omega_{\mu\nu}=\partial_{\mu}u_{\nu}$, etc., which are used in
deriving the covariant chiral kinetic equation in this paper.

\section{Derivation of CCKE}

\label{sec:ccke}Now we start to derive the covariant chiral kinetic
equation from the second line of Eq. (\ref{eq:wig-eq}). To this end
we insert the solution (\ref{eq:1st-solution}) into it, 
\begin{eqnarray}
\nabla_{\mu}[\mathscr{J}_{(0)s}^{\mu}+\mathscr{J}_{(1)s}^{\mu}] & = & \delta(p^{2})\left[p^{\mu}\nabla_{\mu}f_{s}+sQ\frac{1}{p^{2}}\Omega^{\mu\lambda}p_{\lambda}\tilde{F}_{\mu\kappa}p^{\kappa}f_{s}^{\prime}\right.\nonumber \\
 &  & \left.-\frac{s}{2}\tilde{\Omega}^{\mu\lambda}p_{\lambda}(\nabla_{\mu}f_{s}^{\prime})-sQ\frac{1}{p^{2}}\tilde{F}^{\mu\lambda}p_{\lambda}(\nabla_{\mu}f_{s})\right]=0,\label{eq:eom}
\end{eqnarray}
whose derivation is given in Appendix \ref{sec:eom}. We can simplify
$\nabla_{\mu}f_{s}^{\prime}$ as 
\begin{equation}
\nabla_{\mu}f_{s}^{\prime}=\nabla_{\mu}(u_{\nu}\partial_{p}^{\nu}f_{s})=\Omega_{\mu\nu}\partial_{p}^{\nu}f_{s}+u_{\nu}\partial_{p}^{\nu}\nabla_{\mu}f_{s},
\end{equation}
using $f_{s}^{\prime}=df_{s}/dp_{0}=u^{\mu}\partial_{\mu}^{p}f_{s}$,
$\partial_{\mu}u_{\nu}=\Omega_{\mu\nu}$ and $[\partial_{p}^{\nu},\nabla_{\mu}]=0$.
Then Eq. (\ref{eq:eom}) can be rewritten as 
\begin{eqnarray}
\left[p^{\mu}\nabla_{\mu}f_{s}-sQ\frac{1}{p^{2}}\tilde{F}^{\mu\lambda}p_{\lambda}(\nabla_{\mu}f_{s})+sQ\frac{1}{p^{2}}\Omega^{\mu\lambda}p_{\lambda}\tilde{F}_{\mu\kappa}p^{\kappa}u^{\nu}\partial_{\nu}^{p}f_{s}\right.\nonumber \\
\left.-\frac{s}{2}\tilde{\Omega}^{\mu\lambda}p_{\lambda}\Omega_{\mu\nu}\partial_{p}^{\nu}f_{s}-\frac{s}{2}\tilde{\Omega}^{\mu\lambda}p_{\lambda}u_{\nu}\partial_{p}^{\nu}\nabla_{\mu}f_{s}\right]\delta(p^{2}) & = & 0.\label{eq:eom-1}
\end{eqnarray}
Now we try to rewrite the last term $\sim\partial_{p}^{\nu}\nabla_{\mu}f_{s}$.
We assume that the CCKE holds at the momentum integral level, we then
look at the momentum integral of the last term in Eq. (\ref{eq:eom-1})
\begin{eqnarray}
 &  & \int dp_{0}\delta(p^{2})\tilde{\Omega}^{\mu\lambda}p_{\lambda}u_{\nu}\partial_{p}^{\nu}\nabla_{\mu}f_{s}\nonumber \\
 & = & \int dp_{0}\frac{d}{dp_{0}}\left[\delta(p^{2})\tilde{\Omega}^{\mu\lambda}p_{\lambda}\nabla_{\mu}f_{s}\right]-\int dp_{0}\frac{d}{dp_{0}}\left[\delta(p^{2})\tilde{\Omega}^{\mu\lambda}p_{\lambda}\right]\nabla_{\mu}f_{s}\nonumber \\
 & = & -\int dp_{0}\frac{d\delta(p^{2})}{2p_{0}dp_{0}}2p_{0}\tilde{\Omega}^{\mu\lambda}p_{\lambda}\nabla_{\mu}f_{s}-\int dp_{0}\delta(p^{2})\tilde{\Omega}^{\mu\lambda}u_{\lambda}\nabla_{\mu}f_{s}\nonumber \\
 & = & \int dp_{0}\delta(p^{2})\left[\frac{2p_{0}}{p^{2}}\tilde{\Omega}^{\mu\lambda}p_{\lambda}-\tilde{\Omega}^{\mu\lambda}u_{\lambda}\right]\nabla_{\mu}f_{s}
\end{eqnarray}
where we have used in the first equality $d/dp_{0}=u^{\mu}\partial_{\mu}^{p}$
and dropped the surface term. So Eq. (\ref{eq:eom-1}) becomes 
\begin{eqnarray}
\left[\left(p^{\mu}-sQ\frac{1}{p^{2}}\tilde{F}^{\mu\lambda}p_{\lambda}-s\frac{p_{0}}{p^{2}}\tilde{\Omega}^{\mu\lambda}p_{\lambda}+\frac{s}{2}\tilde{\Omega}^{\mu\lambda}u_{\lambda}\right)\nabla_{\mu}f_{s}\right.\nonumber \\
\left.+\left(-\frac{s}{2}\tilde{\Omega}^{\mu\lambda}p_{\lambda}\Omega_{\mu\nu}+sQ\frac{1}{p^{2}}\Omega^{\mu\lambda}p_{\lambda}\tilde{F}_{\mu\kappa}p^{\kappa}u_{\nu}\right)\partial_{p}^{\nu}f_{s}\right]\delta(p^{2}) & = & 0,
\end{eqnarray}
from which we can extract the equations of motion 
\begin{eqnarray}
m_{0}\frac{dx^{\mu}}{d\tau} & = & p^{\mu}-sQ\frac{1}{p^{2}}\tilde{F}^{\mu\lambda}p_{\lambda}-s\frac{p_{0}}{p^{2}}\tilde{\Omega}^{\mu\lambda}p_{\lambda}+\frac{s}{2}\tilde{\Omega}^{\mu\lambda}u_{\lambda},\nonumber \\
m_{0}\frac{dp^{\mu}}{d\tau} & = & QF^{\mu\nu}p_{\nu}+sQ^{2}\frac{1}{4p^{2}}F^{\nu\lambda}\tilde{F}_{\nu\lambda}p^{\mu}+sQ\frac{1}{p^{2}}\Omega_{\nu\lambda}p^{\lambda}\tilde{F}^{\nu\kappa}p_{\kappa}u^{\mu}\nonumber \\
 &  & -\frac{s}{8}\Omega^{\nu\lambda}\tilde{\Omega}_{\nu\lambda}p^{\mu}+\frac{1}{2}sQF^{\mu\nu}\tilde{\Omega}_{\nu\lambda}u^{\lambda}-sQ\frac{p_{0}}{p^{2}}F^{\mu\nu}\tilde{\Omega}_{\nu\lambda}p^{\lambda},\label{eq:ccke-2}
\end{eqnarray}
where $m_{0}$ is an arbitrary mass scale which is irrelevant to the
physics we discuss in this paper and $\tau$ denotes a parameter along
the worldline. In deriving Eq. (\ref{eq:ccke-2}) we have used two
identities $F^{\mu\nu}\tilde{F}_{\nu\lambda}p^{\lambda}=-\frac{1}{4}F^{\rho\sigma}\tilde{F}_{\rho\sigma}p^{\mu}$
and $\Omega^{\mu\nu}\tilde{\Omega}_{\nu\lambda}p^{\lambda}=-\frac{1}{4}\Omega^{\rho\sigma}\tilde{\Omega}_{\rho\sigma}p^{\mu}$
to rewrite two terms of $dp^{\mu}/d\tau$, whose proof is given in
Eq. (\ref{eq:fft-1}). To compare with the previous result in Ref.
\cite{Chen:2012ca}, we rewrite the vorticity part of Eq. (\ref{eq:ccke-2})
in terms of time-like and space-like components of vorticity and field
strength tensor under the conditions in (\ref{eq:static-equil}) 
\begin{eqnarray}
m_{0}\frac{dx^{\mu}}{d\tau} & = & p^{\mu}-sQ\frac{1}{p^{2}}\tilde{F}^{\mu\lambda}p_{\lambda}+s\left(\frac{1}{2}-\frac{p_{0}^{2}}{p^{2}}\right)\omega^{\mu}+s\frac{p_{0}}{p^{2}}(p\cdot\omega)u^{\mu},\nonumber \\
m_{0}\frac{dp^{\mu}}{d\tau} & = & QF^{\mu\nu}p_{\nu}+sQ^{2}\frac{p^{\mu}}{4p^{2}}F^{\nu\lambda}\tilde{F}_{\nu\lambda}\nonumber \\
 &  & +\frac{1}{2}sQ(E\cdot\omega)u^{\mu}-sQ\frac{1}{p^{2}}(p\cdot\omega)(p\cdot E)u^{\mu}+sQ\frac{1}{p^{2}}p_{0}(p\cdot\omega)E^{\mu},\label{eq:ccke-4}
\end{eqnarray}
where $p_{0}\equiv p\cdot u$. The detailed derivation of the vorticity
part of Eq. (\ref{eq:ccke-4}) is given in Appendix \ref{sub:derive-ccke2}.
So the covariant chiral kinetic equation reads 
\begin{equation}
\delta(p^{2})\left(\frac{dx^{\mu}}{d\tau}\partial_{\mu}^{x}f_{s}+\frac{dp^{\mu}}{d\tau}\partial_{\mu}^{p}f_{s}\right)=0,\label{eq:ccke-1}
\end{equation}
with $dx^{\mu}/d\tau$ and $dp^{\mu}/d\tau$ being given by Eq. (\ref{eq:ccke-2})
or (\ref{eq:ccke-4}). 

Note that the covariant chiral kinetic equation (\ref{eq:ccke-4})
is different from Eq. (12) of Ref. \cite{Chen:2012ca} in two places:
(a) there is an additional term $sQp^{-2}(p\cdot\omega)\epsilon^{\mu\nu\rho\sigma}p_{\nu}u_{\rho}B_{\sigma}$
from the last term of $dp^{\mu}/d\tau$ in Eq. (12) of Ref. \cite{Chen:2012ca};
(b) there is a factor 2 in the term $s(p\cdot\omega)(p\cdot u)u^{\mu}/p^{2}$
(the last term of $dx^{\mu}/d\tau$) in Eq. (12) of Ref. \cite{Chen:2012ca}. 

We note that the term in (a) is vanishing when combined with $\partial_{\mu}^{p}f_{s}$
if we use $\partial_{\mu}^{p}f_{s}=u^{\mu}df_{s}/dp_{0}$. The term
in (b) is also vanishing when combined with $\partial_{\mu}^{x}f_{s}$
if we use $\partial_{\mu}^{x}f_{s}=[\partial_{\mu}^{x}(u\cdot p)-\partial_{\mu}^{x}\mu]df_{s}/dp_{0}$
and Eq. (\ref{eq:static-equil}), so it seems that the factor of this
term would be irrelevant. However the coefficient of this term is
essential to obtain the correct energy-momentum tensor from $dx^{\mu}/d\tau$.
We will demonstrate it in great details in the rest part of the paper.

\section{Freedom of adding more terms}

\label{sec:free}Our starting point is the covariant chiral kinetic
equation (\ref{eq:ccke-4}). But the problem with this equation is
that one cannot get the correct energy-momentum tensor from $dx^{\mu}/d\tau$.
One has to add terms to $dx^{\mu}/d\tau$ and $dp^{\mu}/d\tau$ to
achieve this goal. Actually there is a freedom to do so provided the
correct vector and axial vector currents and energy-momentum tensor
are obtained from $dx^{\mu}/d\tau$ after integration over momentum.
In this section, we will give the concrete form of additional terms
allowed by the covariant chiral kinetic equation and conservation
laws. 

We know from Eq. (\ref{eq:ccke-4}) that we need to modify the vorticity
terms. Suppose we add a new term $X^{\mu}$ in linear order of $\omega$
to $dx^{\mu}/d\tau$, which will bring a new term $Y^{\mu}$ to $dp^{\mu}/d\tau$
accordingly in order to keep Eq. (\ref{eq:ccke-1}) hold. So we obtain
the constraint 
\begin{equation}
X^{\sigma}\partial_{\sigma}^{x}f_{s}+Y^{\sigma}\partial_{\sigma}^{p}f_{s}=0.\label{eq:x-y-rel}
\end{equation}
We assume $X^{\mu}$ has the following form 
\begin{eqnarray}
X^{\mu} & = & sC_{1}(p,u)\omega^{\mu}+sC_{2}(p,u)(p\cdot\omega)u^{\mu}\nonumber \\
 &  & +sC_{3}(p,u)(p\cdot\omega)\bar{p}^{\mu},\label{eq:x-mu}
\end{eqnarray}
where $\bar{p}^{\mu}\equiv p^{\mu}-p_{0}u^{\mu}$ and $C_{1,2,3}(p,u)$
are functions of $u^{\mu}$ and $p^{\mu}$ as follows, 
\begin{eqnarray}
C_{1}(p,u) & = & C_{10}+C_{11}\frac{p_{0}^{2}}{p^{2}},\nonumber \\
C_{2}(p,u) & = & C_{20}\frac{p_{0}}{p^{2}}+C_{21}\frac{1}{p_{0}},\nonumber \\
C_{3}(p,u) & = & C_{30}\frac{1}{p^{2}},\label{eq:c0-c1}
\end{eqnarray}
with $\{C_{10},C_{11},C_{20},C_{21},C_{30}\}$ being dimensionless
coefficients to be determined. With $X^{\sigma}$ in Eq. (\ref{eq:x-mu}),
as shown in Appendix \ref{sec:y-vect}, we can solve $Y^{\sigma}$
from Eq. (\ref{eq:x-y-rel}) as, 
\begin{equation}
Y^{\sigma}=-sQ[C_{1}(p,u)(\omega\cdot E)+C_{3}(p,u)(p\cdot\omega)(p\cdot E)]u^{\sigma}+sQ\bar{p}^{\sigma}C_{4}(p,\omega),\label{eq:y-vector}
\end{equation}
where the function $C_{4}(p,\omega)$ is defined by 
\begin{equation}
C_{4}(p,\omega)=C_{40}(\omega\cdot E)\frac{1}{p_{0}}+C_{41}\frac{1}{p^{2}p_{0}}(p\cdot\omega)(p\cdot E),
\end{equation}
with $C_{40}$ and $C_{41}$ being two dimensionless coefficients
to be determined. 

With these terms in (\ref{eq:x-mu},\ref{eq:y-vector}), Eq. (\ref{eq:ccke-4})
is modified to 
\begin{eqnarray}
m_{0}\frac{dx^{\mu}}{d\tau} & = & p^{\mu}-sQ\frac{1}{p^{2}}\tilde{F}^{\mu\lambda}p_{\lambda}\nonumber \\
 &  & +s\left[\frac{1}{2}+C_{10}+(C_{11}-1)\frac{p_{0}^{2}}{p^{2}}\right]\omega^{\mu}\nonumber \\
 &  & +s\left[(C_{20}+1)\frac{p_{0}}{p^{2}}+C_{21}\frac{1}{p_{0}}\right](p\cdot\omega)u^{\mu}\nonumber \\
 &  & +sC_{30}\frac{1}{p^{2}}(p\cdot\omega)\bar{p}^{\mu},\nonumber \\
m_{0}\frac{dp^{\mu}}{d\tau} & = & QF^{\mu\nu}p_{\nu}+sQ^{2}\frac{p^{\mu}}{4p^{2}}F^{\nu\lambda}\tilde{F}_{\nu\lambda}\nonumber \\
 &  & +sQ\left(\frac{1}{2}-C_{10}-C_{11}\frac{p_{0}^{2}}{p^{2}}\right)(\omega\cdot E)u^{\mu}\nonumber \\
 &  & -sQ(C_{30}+1)\frac{1}{p^{2}}(p\cdot\omega)(p\cdot E)u^{\mu}\nonumber \\
 &  & +sQ\bar{p}^{\mu}\left[C_{40}(\omega\cdot E)\frac{1}{p_{0}}+C_{41}\frac{1}{p^{2}p_{0}}(p\cdot\omega)(p\cdot E)\right]\nonumber \\
 &  & +sQ\frac{1}{p^{2}}p_{0}(p\cdot\omega)E^{\mu}.\label{eq:new-ccke}
\end{eqnarray}
The next task is to obtain the constraints for coefficients $\{C_{10},C_{11},C_{20},C_{21},C_{30}\}$
by conservation laws and those for coefficients $\{C_{40},C_{41}\}$
by matching the power (energy rate) to the force for quasi-particles.
Note that all these coefficients in the vorticity terms.

\section{Constraints for coefficients}

\label{sec:constraint}In this section, we give the constraints for
$\{C_{10},C_{11},C_{20},C_{21},C_{30}\}$ by computing the currents
and energy-momentum tensor. The detailed derivation is given in Appendix
\ref{sec:der-curr}.

\subsection{Constraint from currents}

The currents for chiral fermions with chirality $s=\pm1$ are given
by, 
\begin{equation}
j_{s}^{\mu}=m_{0}\int d^{4}p\delta(p^{2})\frac{dx^{\mu}}{d\tau}f_{s}.\label{eq:curr-vort}
\end{equation}

For the electromagnetic term of $dx^{\mu}/d\tau$ in Eq. (\ref{eq:new-ccke}),
we obtain the currents whose derivation is given in Eq. (\ref{eq:j-em}),
\begin{eqnarray}
j_{s}^{\mu}(\mathrm{EM}) & = & -sQ\int d^{4}p\delta(p^{2})\frac{1}{p^{2}}\tilde{F}^{\mu\lambda}p_{\lambda}f_{s}=\xi_{B}^{s}B^{\mu},\label{eq:js-em}
\end{eqnarray}
where we have defined the coefficient (no summation over $s$ is implied)
\begin{eqnarray}
\xi_{B}^{s} & = & \frac{sQ}{4\pi^{2}}\int_{0}^{\infty}dE_{p}\left[f_{\mathrm{FD}}(E_{p}-\mu_{s})-f_{\mathrm{FD}}(E_{p}+\mu_{s})\right].
\end{eqnarray}
Then one can reproduce the chiral magnetic effect for the vector and
axial vector currents as 
\begin{eqnarray}
j^{\mu}(\mathrm{EM}) & = & \sum_{s}j_{s}^{\mu}(\mathrm{EM})=(\xi_{B}^{+}+\xi_{B}^{-})B^{\mu}=\xi_{B}B^{\mu},\nonumber \\
j_{5}^{\mu}(\mathrm{EM}) & == & \sum_{s}sj_{s}^{\mu}(\mathrm{EM})=(\xi_{B}^{+}-\xi_{B}^{-})B^{\mu}=\xi_{B5}B^{\mu},\label{eq:cme-current}
\end{eqnarray}
where $\xi_{B}=Q\mu_{5}/(2\pi^{2})$ and $\xi_{B5}=Q\mu/(2\pi^{2})$
are coefficients in Eqs. (22-23) of Ref. \cite{Gao:2012ix}. 

For the vorticity part of $dx^{\mu}/d\tau$ in Eq. (\ref{eq:new-ccke}),
we obtain the currents induced by the vorticity whose derivation is
shown in Eq. (\ref{eq:der-j-omega}) 
\begin{eqnarray}
j_{s}^{\mu}(\omega) & = & \left(C_{10}-\frac{1}{2}C_{11}+\frac{1}{2}C_{30}+1\right)\xi_{s}\omega^{\mu},\label{eq:curr-vorticity-part}
\end{eqnarray}
where $\xi_{s}$ is defined by (no summation over $s$ is implied)
\begin{eqnarray}
\xi_{s} & \equiv & \int d^{4}p\delta(p^{2})sf_{s}\nonumber \\
 & = & \frac{1}{2\pi^{2}}s\int_{0}^{\infty}dE_{p}E_{p}\left[f_{\mathrm{FD}}(E_{p}-\mu_{s})+f_{\mathrm{FD}}(E_{p}+\mu_{s})\right].
\end{eqnarray}
The vector and axial vector currents in the chiral vortical effect
are given by, 
\begin{eqnarray}
j^{\mu}(\omega) & = & j_{+}^{\mu}(\omega)+j_{-}^{\mu}(\omega)=(\xi_{+}+\xi_{-})\omega^{\mu}=\xi\omega^{\mu},\nonumber \\
j_{5}^{\mu}(\omega) & = & j_{+}^{\mu}(\omega)-j_{-}^{\mu}(\omega)=(\xi_{+}-\xi_{-})\omega^{\mu}=\xi_{5}\omega^{\mu},\label{eq:cve-curr}
\end{eqnarray}
where $\xi=\mu\mu_{5}/\pi^{2}$ and $\xi_{5}=T^{2}/6+(\mu^{2}+\mu_{5}^{2})/(2\pi^{2})$
are coefficients in Eqs. (22-23) of Ref. \cite{Gao:2012ix}. To match
Eq. (\ref{eq:curr-vorticity-part}) with the currents of the chiral
vortical effect in (\ref{eq:cve-curr}), we obtain the first constraint
for $\{C_{10},C_{11},C_{30}\}$, 
\begin{equation}
C_{10}-\frac{1}{2}C_{11}+\frac{1}{2}C_{30}=0.\label{eq:curr}
\end{equation}

\subsection{Constraint from stress tensor}

The energy momentum tensor in the relativistic chiral kinetic theory
can be obtained by 
\begin{equation}
T^{\rho\sigma}=\frac{1}{2}m_{0}\int d^{4}p\delta(p^{2})\sum_{s}\left[p^{\rho}\frac{dx^{\sigma}}{d\tau}f_{s}+p^{\sigma}\frac{dx^{\rho}}{d\tau}f_{s}\right].\label{eq:stress-t}
\end{equation}

First we look at the electromagnetic field part $T^{\rho\sigma}(\mathrm{EM})$,
which we obtain from Eq. (\ref{eq:t-em}), 
\begin{eqnarray}
T^{\rho\sigma}(\mathrm{EM}) & = & \frac{1}{2}Q\xi u^{(\rho}B^{\sigma)},\label{eq:t-em-cve}
\end{eqnarray}
where $\xi$ is the same as the coefficient of $j^{\mu}(\omega)$
in Eq. (\ref{eq:cve-curr}). The energy-momentum tensor (\ref{eq:t-em-cve})
is just the result in Ref. \cite{Gao:2012ix}. 

Now we work on the vorticity part of $T^{\mu\nu}$. Inserting the
last three terms of $dx^{\mu}/d\tau$ in Eq. (\ref{eq:new-ccke})
into Eq. (\ref{eq:stress-t}), we obtain 
\begin{eqnarray}
T^{\rho\sigma}(\omega) & = & \left(\frac{1}{2}C_{10}-\frac{1}{4}C_{11}+\frac{1}{4}C_{30}+\frac{1}{4}C_{20}-\frac{1}{6}C_{21}+\frac{3}{4}\right)n_{5}u^{(\rho}\omega^{\sigma)},\label{eq:emt}
\end{eqnarray}
whose derivation is given in Eq. (\ref{eq:stress-vorticity}). To
match Eq. (\ref{eq:emt}) with the stress tensor of the vorticity
part in Ref. \cite{Gao:2012ix}, $T^{\mu\nu}(\omega)=n_{5}u^{(\rho}\omega^{\sigma)}$
{[}see Eq. (\ref{eq:der-t-vort}) for a derivation{]}, we arrive at
the second constraint for coefficients, 
\begin{eqnarray}
\frac{1}{2}C_{10}-\frac{1}{4}C_{11}+\frac{1}{4}C_{30}+\frac{1}{4}C_{20}-\frac{1}{6}C_{21}+\frac{3}{4} & = & 1.\label{eq:stress}
\end{eqnarray}

Combining Eqs. (\ref{eq:curr},\ref{eq:stress}) we obtain two independent
constraints 
\begin{eqnarray}
C_{10}-\frac{1}{2}C_{11}+\frac{1}{2}C_{30} & = & 0,\nonumber \\
C_{20}-\frac{2}{3}C_{21} & = & 1.\label{eq:c0-c2}
\end{eqnarray}
We see that $C_{20}$ and $C_{21}$ cannot all be zero.

\section{Chiral kinetic equation: from 4D to 3D}

\label{sec:4d-to-3d}From the covariant chiral kinetic equation (\ref{eq:ccke-1})
with $dx^{\mu}/d\tau$ and $dp^{\mu}/d\tau$ given by Eq. (\ref{eq:new-ccke}),
we can obtain its 3D version by integrating over $p_{0}$, 
\begin{eqnarray}
I & = & \int dp_{0}\delta(p^{2})\left[\frac{dx^{\sigma}}{d\tau}\partial_{\sigma}^{x}f_{s}+\frac{dp^{\sigma}}{d\tau}\partial_{\sigma}^{p}f_{s}\right]\nonumber \\
 & = & I_{x0}+I_{p0}+I_{x}+I_{p},\label{eq:ix-ip}
\end{eqnarray}
where we work in the co-moving frame with $u^{\mu}=(1,0)$, and $I_{x0}$,
$I_{p0}$, $I_{x}$ and $I_{p}$ are from $(dx^{0}/d\tau)\partial_{0}^{x}f_{s}$,
$(dp^{0}/d\tau)\partial_{0}^{p}f_{s}$, $(dx^{i}/d\tau)\partial_{i}^{x}f_{s}$,
$(dp^{i}/d\tau)\partial_{i}^{p}f_{s}$ respectively. We evaluate each
term in Appendix \ref{sec:derivation-ccke}. In evaluation of these
terms we imply an additional integration over $\mathbf{p}$, i.e.$\int d^{3}p$,
so we can drop complete derivative terms in $\mathbf{p}$ (or $E_{p}$)
which are vanishing at the boundary. 

The results for $I_{x0}$ are derived in Eqs. (\ref{eq:ix0}-\ref{eq:ix0-vort}), 

\begin{eqnarray}
I_{x0}(0) & = & \frac{1}{(2\pi)^{3}}\partial_{0}^{x}\left[f_{\mathrm{FD}}(E_{p}-\mu_{s})-f_{\mathrm{FD}}(E_{p}+\mu_{s})\right],\nonumber \\
I_{x0}(\mathrm{EM}) & = & \frac{1}{(2\pi)^{3}}\left\{ sQ(\mathbf{p}\cdot\mathbf{B})\frac{1}{2E_{p}^{3}}\partial_{0}^{x}\left[f_{\mathrm{FD}}(E_{p}-\mu_{s})+f_{\mathrm{FD}}(E_{p}+\mu_{s})\right]\right.\nonumber \\
 &  & \left.-sQ(\mathbf{p}\cdot\mathbf{B})\frac{1}{2E_{p}^{2}}\frac{d}{dE_{p}}\partial_{0}^{x}\left[f_{\mathrm{FD}}(E_{p}-\mu_{s})+f_{\mathrm{FD}}(E_{p}+\mu_{s})\right]\right\} ,\nonumber \\
 & \equiv & I_{x0}^{1}(\mathrm{EM})+I_{x0}^{2}(\mathrm{EM})\nonumber \\
I_{x0}(\omega) & = & \frac{1}{(2\pi)^{3}}s(C_{20}-C_{21}+1)(\mathbf{p}\cdot\boldsymbol{\omega})\frac{1}{E_{p}^{2}}\partial_{0}^{x}\left[f_{\mathrm{FD}}(E_{p}-\mu_{s})-f_{\mathrm{FD}}(E_{p}+\mu_{s})\right],\label{eq:ix0-em1}
\end{eqnarray}
where we have labeled two terms of $I_{x0}(\mathrm{EM})$ as $I_{x0}^{1}(\mathrm{EM})$
and $I_{x0}^{2}(\mathrm{EM})$ for later use. The results for $I_{x}$
are derived in Eqs. (\ref{eq:ix}-\ref{eq:ix-vort}), 
\begin{eqnarray}
I_{x}(0) & = & \int dp_{0}\delta(p^{2})p^{i}\partial_{i}^{x}f_{s}\nonumber \\
 & = & \frac{\mathbf{p}_{i}}{E_{p}}\frac{1}{(2\pi)^{3}}\partial_{i}^{x}\left[f_{\mathrm{FD}}(E_{p}-\mu_{s})+f_{\mathrm{FD}}(E_{p}+\mu_{s})\right],\nonumber \\
I_{x}(\mathrm{EM}) & = & \frac{1}{(2\pi)^{3}}sQ\mathbf{B}_{i}\frac{1}{2E_{p}^{2}}\partial_{i}^{x}\left[f_{\mathrm{FD}}(E_{p}-\mu_{s})-f_{\mathrm{FD}}(E_{p}+\mu_{s})\right]\nonumber \\
 &  & -\frac{1}{(2\pi)^{3}}sQ(\mathbf{p}\times\mathbf{E})_{i}\frac{1}{2E_{p}^{3}}\partial_{i}^{x}\left[f_{\mathrm{FD}}(E_{p}-\mu_{s})+f_{\mathrm{FD}}(E_{p}+\mu_{s})\right]\nonumber \\
 &  & +\frac{1}{(2\pi)^{3}}sQ(\mathbf{p}\times\mathbf{E})_{i}\frac{1}{2E_{p}^{2}}\frac{d}{dE_{p}}\partial_{i}^{x}\left[f_{\mathrm{FD}}(E_{p}-\mu_{s})+f_{\mathrm{FD}}(E_{p}+\mu_{s})\right],\nonumber \\
 & \equiv & I_{x}^{1}(\mathrm{EM})+I_{x}^{2}(\mathrm{EM})+I_{x}^{3}(\mathrm{EM})\nonumber \\
I_{x}(\omega) & = & \frac{1}{(2\pi)^{3}}s\left(1+C_{10}-\frac{1}{2}C_{11}\right)\frac{1}{E_{p}}\boldsymbol{\omega}^{i}\partial_{i}^{x}\left[f_{F}(E_{p}-\mu_{s})+f_{F}(E_{p}+\mu_{s})\right]\nonumber \\
 &  & +\frac{1}{(2\pi)^{3}}sC_{30}(\mathbf{p}\cdot\boldsymbol{\mathbf{\omega}})\mathbf{p}_{i}\frac{3}{2E_{p}^{3}}\partial_{i}^{x}\left[f_{F}(E_{p}-\mu_{s})+f_{F}(E_{p}+\mu_{s})\right],\label{eq:ix-em1}
\end{eqnarray}
where we have labeled three terms of $I_{x}(\mathrm{EM})$ as $I_{x}^{1}(\mathrm{EM})$,
$I_{x}^{2}(\mathrm{EM})$ and $I_{x}^{3}(\mathrm{EM})$ for later
use. 

The results for $I_{p0}$ and $I_{p}$ are derived in Eqs. (\ref{eq:ip0-em},\ref{eq:ip0},\ref{eq:ip-em},\ref{eq:ip-vort}),
\begin{eqnarray}
I_{p0}(\mathrm{EM}) & = & \frac{1}{(2\pi)^{3}}Q(\mathbf{p}\cdot\mathbf{E})\frac{1}{E_{p}}\frac{d}{dE_{p}}\left[f_{\mathrm{FD}}(E_{p}-\mu_{s})-f_{\mathrm{FD}}(E_{p}+\mu_{s})\right]\nonumber \\
 &  & +\frac{1}{(2\pi)^{3}}sQ^{2}(\mathbf{E}\cdot\mathbf{B})\frac{1}{2E_{p}^{2}}\frac{d}{dE_{p}}\left[f_{\mathrm{FD}}(E_{p}-\mu_{s})+f_{\mathrm{FD}}(E_{p}+\mu_{s})\right],\nonumber \\
I_{p0}(\omega) & = & \frac{1}{(2\pi)^{3}}sQ(C_{30}+1)\left[-\frac{1}{2E_{p}}(\mathbf{E}\cdot\boldsymbol{\omega})+\frac{3}{2E_{p}^{3}}(\mathbf{p}\cdot\boldsymbol{\omega})(\mathbf{p}\cdot\mathbf{E})\right]\nonumber \\
 &  & \frac{d}{dE_{p}}\left[f_{\mathrm{FD}}(E_{p}-\mu_{s})-f_{\mathrm{FD}}(E_{p}+\mu_{s})\right],\label{eq:ip0-1}\\
I_{p}(\mathrm{EM}) & = & \frac{1}{(2\pi)^{3}}\left\{ Q\mathbf{E}_{i}\partial_{i}^{p}\left[f_{\mathrm{FD}}(E_{p}-\mu_{s})-f_{\mathrm{FD}}(E_{p}+\mu_{s})\right]\right.\nonumber \\
 &  & \left.+\left[Q\frac{\mathbf{p}}{E_{p}}\times\mathbf{B}+sQ^{2}(\mathbf{E}\cdot\mathbf{B})\frac{\mathbf{p}}{E_{p}^{3}}\right]_{i}\partial_{i}^{p}\left[f_{\mathrm{FD}}(E_{p}-\mu_{s})+f_{\mathrm{FD}}(E_{p}+\mu_{s})\right]\right\} ,\nonumber \\
I_{p}(\omega) & = & \frac{1}{(2\pi)^{3}}sQ\left[\frac{1}{E_{p}^{2}}(\mathbf{p}\cdot\boldsymbol{\omega})\mathbf{E}_{i}-C_{40}\frac{1}{E_{p}^{2}}(\boldsymbol{\omega}\cdot\mathbf{E})\mathbf{p}_{i}-C_{41}\frac{2}{E_{p}^{4}}(\mathbf{p}\cdot\boldsymbol{\omega})(\mathbf{p}\cdot\mathbf{E})\mathbf{p}_{i}\right]\nonumber \\
 &  & \partial_{i}^{p}\left[f_{\mathrm{FD}}(E_{p}-\mu_{s})-f_{\mathrm{FD}}(E_{p}+\mu_{s})\right],\label{eq:ip-1}
\end{eqnarray}

We now extract $d\mathbf{p}/d\tau$ from Eqs. (\ref{eq:ip0-1},\ref{eq:ip-1})
for particles (it is similar for anti-particles). For an on-shell
particle the energy is not an independent phase space variable, its
rate $dE_{p}/d\tau$ from $I_{p0}$ can be determined by (see, e.g.,
Ref. \cite{Groot:Maxwell}) 
\begin{equation}
\frac{dE_{p}}{d\tau}=\frac{1}{E_{p}}\mathbf{p}\cdot\frac{d\mathbf{p}}{d\tau}.\label{eq:e-p-rel}
\end{equation}
So in derivation of the 3D chiral kinetic equation from the 4D one,
the $p_{0}$ degree of freedom is fixed and is not a kinematic variable
in the 3D equation. 

First we look at $I_{p}(\mathrm{EM})$ in Eq. (\ref{eq:ip0-1}) from
which we can obtain the energy rate from the electromagnetic field
\begin{eqnarray}
\frac{dE_{p}}{d\tau}(\mathrm{EM}) & = & Q(\mathbf{p}\cdot\mathbf{E})\frac{1}{E_{p}}+sQ^{2}(\mathbf{E}\cdot\mathbf{B})\frac{1}{2E_{p}^{2}}.\label{eq:de-dtau-em}
\end{eqnarray}
Let us compare it with $d\mathbf{p}/d\tau$ extracted from $I_{p}(\mathrm{EM})$
in Eq. (\ref{eq:ip-1}), 
\begin{eqnarray}
\frac{d\mathbf{p}}{d\tau}(\mathrm{EM}) & = & Q\left(\mathbf{E}+\frac{\mathbf{p}}{E_{p}}\times\mathbf{B}\right)+sQ^{2}(\mathbf{E}\cdot\mathbf{B})\mathbf{p}\frac{1}{E_{p}^{3}}.\label{eq:dp-dtau-em}
\end{eqnarray}
But Eqs. (\ref{eq:de-dtau-em},\ref{eq:dp-dtau-em}) do not satisfy
Eq. (\ref{eq:e-p-rel}) due to a factor 2 difference in the $\mathbf{E}\cdot\mathbf{B}$
term. The solution to this problem is to define $d\mathbf{p}/d\tau$
in such a way that Eq. (\ref{eq:e-p-rel}) is satisfied 
\begin{equation}
\frac{d\mathbf{p}}{d\tau}(\mathrm{EM})=Q\left(\mathbf{E}+\frac{\mathbf{p}}{E_{p}}\times\mathbf{B}\right)+sQ^{2}(\mathbf{E}\cdot\mathbf{B})\mathbf{p}\frac{1}{2E_{p}^{3}},\label{eq:corr-dp-dtau-em}
\end{equation}
where the last term is only one-half of that in Eq. (\ref{eq:dp-dtau-em}).
Note that our current solution of $d\mathbf{p}/d\tau$ is based on
the first order solution of the Wigner function which is linear in
the electromagnetic field except the anomaly term. The other half
of the anomaly term can be grouped with other second order terms of
the electromagnetic fields which belong to the higher order solutions.
This is beyond the scope of our current paper and for a future study. 

In the same way, the energy rate from the vorticity can be obtained
from $I_{p0}(\omega)$ in Eq. (\ref{eq:ip0-1}), 
\begin{eqnarray}
\frac{dE_{p}}{d\tau}(\omega) & = & -sQ(C_{30}+1)\frac{1}{2E_{p}}(\mathbf{E}\cdot\boldsymbol{\omega})\nonumber \\
 &  & +sQ(C_{30}+1)\frac{3}{2E_{p}^{3}}(\mathbf{p}\cdot\boldsymbol{\omega})(\mathbf{p}\cdot\mathbf{E}).\label{eq:de-dtau-vort}
\end{eqnarray}
We can compare it with $d\mathbf{p}/d\tau$ extracted from $I_{p}(\omega)$
in Eq. (\ref{eq:ip-1}), 
\begin{eqnarray}
\frac{d\mathbf{p}}{d\tau}(\omega) & = & sQ\frac{1}{E_{p}^{2}}\left[(\mathbf{p}\cdot\boldsymbol{\omega})\mathbf{E}-C_{40}(\boldsymbol{\omega}\cdot\mathbf{E})\mathbf{p}\right.\nonumber \\
 &  & \left.-2C_{41}\frac{1}{E_{p}^{2}}(\mathbf{p}\cdot\boldsymbol{\omega})(\mathbf{p}\cdot\mathbf{E})\mathbf{p}\right],\label{eq:dp-dtau-vort}
\end{eqnarray}
Applying Eq. (\ref{eq:e-p-rel}) for Eqs. (\ref{eq:de-dtau-vort},\ref{eq:dp-dtau-vort}),
we obtain following constraints for coefficients 
\begin{eqnarray}
C_{40} & = & \frac{1}{2}(C_{30}+1),\nonumber \\
C_{41} & = & \frac{1}{2}-\frac{3}{4}(C_{30}+1).\label{eq:c40-41}
\end{eqnarray}
If we choose $C_{30}=0$, we obtain $C_{40}=1/2$ and $C_{41}=-1/4$.
Then from Eqs. (\ref{eq:dp-dtau-em},\ref{eq:dp-dtau-vort}) we have
\begin{eqnarray}
\frac{d\mathbf{p}}{d\tau} & = & Q\left(\mathbf{E}+\frac{\mathbf{p}}{|\mathbf{p}|}\times\mathbf{B}\right)+sQ^{2}(\mathbf{E}\cdot\mathbf{B})\boldsymbol{\Omega}\nonumber \\
 &  & +sQ\frac{1}{|\mathbf{p}|^{2}}\left[(\mathbf{p}\cdot\boldsymbol{\omega})\mathbf{E}-\frac{1}{2}(\boldsymbol{\omega}\cdot\mathbf{E})\mathbf{p}+\frac{1}{2|\mathbf{p}|^{2}}(\mathbf{p}\cdot\boldsymbol{\omega})(\mathbf{p}\cdot\mathbf{E})\mathbf{p}\right],\label{eq:ccke-dp}
\end{eqnarray}
where $\boldsymbol{\Omega}=\mathbf{p}/(2|\mathbf{p}|^{3})$ is the
Berry curvature in 3-momentum space. Equation (\ref{eq:ccke-dp})
differs from Eq. (22) of Ref. \cite{Chen:2012ca} only by an additional
term perpendicular to $\mathbf{p}$: $sQ|\mathbf{p}|^{-4}(\mathbf{p}\cdot\boldsymbol{\omega})\mathbf{E}_{T}$
with $\mathbf{E}_{T}\equiv\mathbf{E}-(\hat{\mathbf{p}}\cdot\mathbf{E})\hat{\mathbf{p}}$.
This additional term is allowed by the energy and momentum constraint
(\ref{eq:e-p-rel}). 

Now we try to extract $dx_{0}/d\tau$ and $d\mathbf{x}/d\tau$ from
$I_{x0}$ and $I_{x}$ in Eqs. (\ref{eq:ix0-em1},\ref{eq:ix-em1}).
We can combine Eqs. (\ref{eq:ix0-em1},\ref{eq:ix-em1}) to obtain
$dx_{0}/d\tau$ and $d\mathbf{x}/d\tau$ for fermions (one can also
obtain results for anti-fermions similarly)

\begin{eqnarray}
\frac{dx_{0}}{d\tau} & = & 1+\mathscr{C}_{B}sQ(\boldsymbol{\Omega}\cdot\mathbf{B})+\left(4-\frac{2}{3}C_{21}\right)s|\mathbf{p}|(\boldsymbol{\Omega}\cdot\boldsymbol{\omega}),\nonumber \\
\frac{d\mathbf{x}}{d\tau} & = & \hat{\mathbf{p}}+sQ\mathbf{B}(\hat{\mathbf{p}}\cdot\boldsymbol{\Omega})+\mathscr{C}_{E}sQ(\mathbf{E}\times\boldsymbol{\Omega})\nonumber \\
 &  & +s\left(1-\frac{1}{2}C_{30}\right)\frac{\boldsymbol{\omega}}{|\mathbf{p}|}+3sC_{30}(\boldsymbol{\Omega}\cdot\boldsymbol{\mathbf{\omega}})\mathbf{p},\label{eq:ccke-dxdtau}
\end{eqnarray}
where we have used the constraint (\ref{eq:c0-c2}). Here $\mathscr{C}_{B}$
and $\mathscr{C}_{E}$ are prefactors of the $sQ(\boldsymbol{\Omega}\cdot\mathbf{B})$,
$sQ(\mathbf{E}\times\boldsymbol{\Omega})$ terms, respectively, that
need to be determined. The freedom to choose $\mathscr{C}_{B}$ and
$\mathscr{C}_{E}$ is because the integration over $\mathbf{p}$ of
$I_{x0}^{1}(\mathrm{EM}),I_{x0}^{2}(\mathrm{EM}),I_{x}^{2}(\mathrm{EM}),I_{x}^{3}(\mathrm{EM})$
in Eqs. (\ref{eq:ix0-em1},\ref{eq:ix-em1}) are all vanishing, we
can make choices as to keep or drop them following some physical reasons.
So $\mathscr{C}_{B}$ and $\mathscr{C}_{E}$ can be either 1 or 2.
The value 2 is due to the fact that the $\mathbf{p}$ integrals of
$I_{x0}^{1}(\mathrm{EM})$ and $I_{x0}^{2}(\mathrm{EM})$ are equal
and so are those of $I_{x}^{2}(\mathrm{EM})$ and $I_{x}^{3}(\mathrm{EM})$.
We can set $\mathscr{C}_{E}$ to 1 to match the previous result. 

We can now combine Eq. (\ref{eq:corr-dp-dtau-em}), (\ref{eq:dp-dtau-vort})
and (\ref{eq:ccke-dxdtau}) to obtain the 3D chiral kinetic equation,
\begin{eqnarray}
\frac{dx_{0}}{d\tau} & = & 1+\mathscr{C}_{B}sQ(\boldsymbol{\Omega}\cdot\mathbf{B})+\left(4-\frac{2}{3}C_{21}\right)s|\mathbf{p}|(\boldsymbol{\Omega}\cdot\boldsymbol{\omega}),\nonumber \\
\frac{d\mathbf{x}}{d\tau} & = & \hat{\mathbf{p}}+sQ\mathbf{B}(\hat{\mathbf{p}}\cdot\boldsymbol{\Omega})+\mathscr{C}_{E}sQ(\mathbf{E}\times\boldsymbol{\Omega})\nonumber \\
 &  & +s\left(1-\frac{1}{2}C_{30}\right)\frac{\boldsymbol{\omega}}{|\mathbf{p}|}+3C_{30}s(\boldsymbol{\Omega}\cdot\boldsymbol{\mathbf{\omega}})\mathbf{p},\nonumber \\
\frac{d\mathbf{p}}{d\tau} & = & Q\left(\mathbf{E}+\frac{\mathbf{p}}{|\mathbf{p}|}\times\mathbf{B}\right)+sQ^{2}(\mathbf{E}\cdot\mathbf{B})\frac{\mathbf{p}}{2|\mathbf{p}|^{3}}\nonumber \\
 &  & +sQ\frac{1}{|\mathbf{p}|^{2}}\left[(\mathbf{p}\cdot\boldsymbol{\omega})\mathbf{E}-\frac{1}{2}(C_{30}+1)(\boldsymbol{\omega}\cdot\mathbf{E})\mathbf{p}\right.\nonumber \\
 &  & \left.+\frac{1}{2}(1+3C_{30})\frac{1}{|\mathbf{p}|^{2}}(\mathbf{p}\cdot\boldsymbol{\omega})(\mathbf{p}\cdot\mathbf{E})\mathbf{p}\right].\label{eq:3d-cke}
\end{eqnarray}
We see that the chiral kinetic equation (\ref{eq:3d-cke}) is not
uniquely determined due to a freedom to choose the coefficients $\{C_{21},C_{30},\mathscr{C}_{B},\mathscr{C}_{E}\}$
where $\mathscr{C}_{B}$ and $\mathscr{C}_{E}$ can be set to 1 or
2. If we choose $\{C_{21},C_{30},\mathscr{C}_{B},\mathscr{C}_{E}\}=\{0,0,1,1\}$,
Eq. (\ref{eq:3d-cke}) reproduces the result of Ref. \cite{Chen:2012ca}
except an additional term perpendicular to $\mathbf{p}$ in $d\mathbf{p}/d\tau$
which is allowed by the energy and momentum constraint (\ref{eq:e-p-rel}).
Another possible choice of $C_{30}$ is $C_{30}=2/3$. In this case
the vorticity terms in $d\mathbf{x}/d\tau$ in Eq. (\ref{eq:3d-cke})
read 
\begin{equation}
\frac{d\mathbf{x}}{d\tau}(\omega)=\frac{s}{|\mathbf{p}|}(\hat{\mathbf{p}}\cdot\boldsymbol{\mathbf{\omega}})\hat{\mathbf{p}}+\frac{s}{|\mathbf{p}|}\frac{2}{3}\boldsymbol{\omega}.\label{eq:dx-dtau-omega}
\end{equation}
When calculating the vorticity contribution to the current from $d\mathbf{x}/d\tau$
by integration over $\mathbf{p}$, one can verify that the first term
of (\ref{eq:dx-dtau-omega}) contributes to 1/3 of the chiral vortical
effect while the second term contributes to the rest 2/3. In comparison
to the result of Ref. \cite{Chen:2014cla,Kharzeev:2016sut}, the 1/3
contribution corresponds to that from the spin-vorticity coupling
energy, while the rest 2/3 contribution corresponds to that from the
magnetization current. 

We can show that the $sQ(\boldsymbol{\Omega}\cdot\mathbf{B})$ term
in $dx_{0}/d\tau$ in Eq. (\ref{eq:3d-cke}) can be absorbed into
the distribution function as the magnetic moment energy $\Delta E_{p}^{m}=sQ\frac{1}{2E_{p}^{2}}(\mathbf{p}\cdot\mathbf{B})$
\cite{Gao:2015zka}. Similarly the term $2s|\mathbf{p}|(\boldsymbol{\Omega}\cdot\boldsymbol{\omega})$
in $dx_{0}/d\tau$ in Eq. (\ref{eq:3d-cke}) can also be absorbed
into the distribution function as the spin-vorticity coupling energy
$\Delta E_{p}^{\omega}=s\frac{1}{2}(\hat{\mathbf{p}}\cdot\boldsymbol{\omega})$
\cite{Gao:2015zka}. But once including the energy corrections into
the distribution functions, $d\mathbf{x}/d\tau$ has to be modified
in order to reproduce the CME and CVE, which has been done in Ref.
\cite{Chen:2014cla,Kharzeev:2016sut}. 

We should point out that our current approach is based on perturbation
in which the zero-th order Wigner functions involve the Fermi-Dirac
distribution functions for free and massless fermions. If the zero-th
order distributions are modified by including e.g. energy corrections
from the magnetic moment and spin-vorticity coupling of fermions,
our current perturbation method breaks down. Another key assumption
of our approach is the static equilibrium conditions under which we
obtain the vector and axial vector components of Wigner functions
up to the first order. Due to these two assumptions, we cannot fix
the coefficients $\{C_{21},C_{30},\mathscr{C}_{B},\mathscr{C}_{E}\}$
in the 3D chiral kinetic equation. To go beyond our current approach,
one needs a new and consistent way of building up the 3D chiral kinetic
equation from the CCKE or directly from covariant Wigner equations.

\section{Summary}

The chiral kinetic equation is an important aspect of chiral fermions,
which is closely related to the Berry phase and monopole in momentum
space. We derive the covariant chiral kinetic equation (CCKE) in the
4-dimensional Wigner function approach using an improved perturbative
method under the static equilibrium conditions. The chiral kinetic
equation in 3-dimensions can be obtained by integration over the time
component of the 4-momentum from the CCKE. In the 3-dimensional equation,
due to the on-shell condition, $p_{0}$ degree of freedom is removed
and is not a phase space variable. There is a freedom to add more
terms into the CCKE whose coefficients can be constrained by conservation
laws. Moreover, in the 3-dimensional equation, the coefficients of
the $sQ(\boldsymbol{\Omega}\cdot\mathbf{B})$ term of $dx_{0}/d\tau$
and the $sQ(\mathbf{E}\times\boldsymbol{\Omega})$ term of $d\mathbf{x}/d\tau$
cannot be fixed in the current formalism where $\boldsymbol{\Omega}$
is the Berry curvature in 3-momentum space. Therefore the 3-dimensional
chiral kinetic equation is not uniquely determined up to some coefficients
in our current approach. By one set of coefficients we can recover
the the 3-dimensional chiral kinetic equation derived in Ref. \cite{Chen:2012ca}
except a transverse electric field term allowed by the energy and
momentum constraint. To go beyond our current approach, one needs
a new and consistent way of building up the 3D chiral kinetic equation
from the CCKE or directly from covariant Wigner equations. 

Acknowledgment. QW thanks J.F. Liao, M. Stephanov, H.-U. Yee and Y.
Yin for helpful discussions. QW is supported in part by the Major
State Basic Research Development Program (MSBRD) in China under the
Grant No. 2015CB856902 and 2014CB845402 and by the National Natural
Science Foundation of China (NSFC) under the Grant No. 11535012. JHG
is supported in part by the Major State Basic Research Development
Program in China under the Grant No. 2014CB845406, the National Natural
Science Foundation of China under the Grant No. 11475104 and the Natural
Science Foundation of Shandong Province under the Grant No. JQ201601. 

\appendix

\section{Derivation of Eq. (\ref{eq:eom})}

\label{sec:eom}In this appendix, we evaluate $\nabla_{\mu}\mathscr{J}_{(1)s}^{\mu}$
with $\mathscr{J}_{(1)s}^{\mu}$ given in Eq. (\ref{eq:1st-solution}).
The final result is Eq. (\ref{eq:eom}). First we list some useful
formula for the field strengths and vorticity tensors, 
\begin{eqnarray}
F^{\mu\nu} & = & E^{\mu}u^{\nu}-E^{\nu}u^{\mu}+\epsilon^{\mu\nu\rho\sigma}u_{\rho}B_{\sigma},\nonumber \\
\tilde{F}^{\mu\nu} & = & B^{\mu}u^{\nu}-B^{\nu}u^{\mu}+\epsilon^{\mu\nu\rho\sigma}E_{\rho}u_{\sigma},\nonumber \\
\Omega^{\mu\nu} & = & \varepsilon^{\mu}u^{\nu}-\varepsilon^{\nu}u^{\mu}+\epsilon^{\mu\nu\rho\sigma}u_{\rho}\omega_{\sigma},\nonumber \\
\tilde{\Omega}^{\mu\nu} & = & \omega^{\mu}u^{\nu}-\omega^{\nu}u^{\mu}+\epsilon^{\mu\nu\rho\sigma}\varepsilon_{\rho}u_{\sigma},\nonumber \\
\Omega^{\mu\nu} & = & \frac{1}{2}(\partial^{\mu}u^{\nu}-\partial^{\nu}u^{\mu})=-\frac{1}{2}\epsilon^{\mu\nu\rho\sigma}\tilde{\Omega}_{\rho\sigma},\nonumber \\
\tilde{\Omega}^{\mu\nu} & = & \frac{1}{2}\epsilon^{\mu\nu\rho\sigma}\Omega_{\rho\sigma}.
\end{eqnarray}
We evaluate the vorticity and electromagnetic field terms as follows,
\begin{eqnarray}
\nabla_{\mu}[\tilde{\Omega}^{\mu\lambda}p_{\lambda}f_{s}^{\prime}\delta(p^{2})] & = & \tilde{\Omega}^{\mu\lambda}\nabla_{\mu}[p_{\lambda}f_{s}^{\prime}\delta(p^{2})]\nonumber \\
 & = & -Q\tilde{\Omega}^{\mu\lambda}F_{\mu\lambda}f_{s}^{\prime}\delta(p^{2})+\tilde{\Omega}^{\mu\lambda}p_{\lambda}(\nabla_{\mu}f_{s}^{\prime})\delta(p^{2})+2Q\frac{1}{p^{2}}\tilde{\Omega}^{\mu\lambda}p_{\lambda}F_{\mu\kappa}p^{\kappa}f_{s}^{\prime}\delta(p^{2})\nonumber \\
 & = & Q\frac{1}{p^{2}}\left[2\tilde{\Omega}^{\mu\lambda}p_{\lambda}F_{\mu\kappa}p^{\kappa}-p^{2}\tilde{\Omega}^{\mu\lambda}F_{\mu\lambda}\right]f_{s}^{\prime}\delta(p^{2})+\tilde{\Omega}^{\mu\lambda}p_{\lambda}(\nabla_{\mu}f_{s}^{\prime})\delta(p^{2})\nonumber \\
 & = & -2Q\frac{1}{p^{2}}\Omega^{\mu\lambda}p_{\lambda}\tilde{F}_{\mu\kappa}p^{\kappa}f_{s}^{\prime}\delta(p^{2})+\tilde{\Omega}^{\mu\lambda}p_{\lambda}(\nabla_{\mu}f_{s}^{\prime})\delta(p^{2}),\\
\nabla_{\mu}\left[\frac{1}{p^{2}}\tilde{F}^{\mu\lambda}p_{\lambda}f_{s}\delta(p^{2})\right] & = & 2Q\frac{1}{p^{4}}\tilde{F}^{\mu\lambda}p_{\lambda}F_{\mu\nu}p^{\nu}f_{s}\delta(p^{2})-Q\frac{1}{p^{2}}\tilde{F}^{\mu\lambda}F_{\mu\lambda}f_{s}\delta(p^{2})\nonumber \\
 &  & +\frac{1}{p^{2}}\tilde{F}^{\mu\lambda}p_{\lambda}(\nabla_{\mu}f_{s})\delta(p^{2})+2Q\frac{1}{p^{4}}\tilde{F}^{\mu\lambda}p_{\lambda}F_{\mu\nu}p^{\nu}f_{s}\delta(p^{2})\nonumber \\
 & = & \frac{1}{p^{2}}\tilde{F}^{\mu\lambda}p_{\lambda}(\nabla_{\mu}f_{s})\delta(p^{2})+Q\frac{1}{p^{4}}\left[4\tilde{F}^{\mu\lambda}p_{\lambda}F_{\mu\nu}p^{\nu}-p^{2}\tilde{F}^{\mu\lambda}F_{\mu\lambda}\right]f_{s}\delta(p^{2})\nonumber \\
 & = & \frac{1}{p^{2}}\tilde{F}^{\mu\lambda}p_{\lambda}(\nabla_{\mu}f_{s})\delta(p^{2}),
\end{eqnarray}
where we have used $\nabla_{\mu}\tilde{\Omega}^{\mu\lambda}=\partial_{\mu}\tilde{\Omega}^{\mu\lambda}=0$
under static-equilibrium conditions and $\nabla_{\mu}\tilde{F}^{\mu\lambda}=0$
for constant electromagnetic field. We also used following formula,
\begin{eqnarray}
\nabla_{\mu}p_{\lambda} & = & -QF_{\mu\lambda},\nonumber \\
\nabla_{\mu}\delta(p^{2}) & = & 2QF_{\mu\nu}p^{\nu}\frac{1}{p^{2}}\delta(p^{2}),\nonumber \\
\nabla_{\mu}\frac{1}{p^{2}} & = & 2QF_{\mu\nu}p^{\nu}\frac{1}{p^{4}},
\end{eqnarray}
and 
\begin{eqnarray}
\tilde{\Omega}^{\mu\lambda}p_{\lambda}F_{\mu\kappa}p^{\kappa}+\Omega^{\mu\lambda}p_{\lambda}\tilde{F}_{\mu\kappa}p^{\kappa} & = & \frac{1}{2}p^{2}\tilde{\Omega}^{\rho\sigma}F_{\rho\sigma},\nonumber \\
4F^{\mu\lambda}p_{\lambda}\tilde{F}_{\mu\nu}p^{\nu}-p^{2}F_{\rho\sigma}\tilde{F}^{\rho\sigma} & = & 0,\nonumber \\
4\Omega^{\mu\lambda}p_{\lambda}\tilde{\Omega}_{\mu\nu}p^{\nu}-p^{2}\Omega^{\rho\sigma}\tilde{\Omega}_{\rho\sigma} & = & 0.\label{eq:id-1}
\end{eqnarray}

We can prove the first identity of (\ref{eq:id-1}) by observing 
\begin{eqnarray*}
\tilde{\Omega}^{\mu\nu}F_{\nu\lambda}p^{\lambda} & = & -\omega^{\mu}(p\cdot E)-(u\cdot p)(\omega\cdot E)u^{\mu},\\
\Omega^{\mu\nu}\tilde{F}_{\nu\lambda}p^{\lambda} & = & (p\cdot\omega)E^{\mu}+(u\cdot p)(\omega\cdot E)u^{\mu}-(\omega\cdot E)p^{\mu},\\
\tilde{\Omega}^{\mu\nu}F_{\mu\nu} & = & 2(\omega\cdot E),
\end{eqnarray*}
where we have used $\varepsilon^{\mu}=\Omega^{\mu\nu}u_{\nu}=u_{\nu}\partial^{\mu}u^{\nu}=0$
and $\epsilon_{\nu\lambda\rho\sigma}\omega^{\nu}u^{\rho}B^{\sigma}=0$
and 
\begin{eqnarray}
-\epsilon^{\nu\mu\alpha\beta}\epsilon_{\nu\lambda\rho\sigma}u_{\alpha}\omega_{\beta}E^{\rho}u^{\sigma}p^{\lambda} & = & (p\cdot\omega)E^{\mu}+(u\cdot p)(\omega\cdot E)u^{\mu}-(\omega\cdot E)p^{\mu}.
\end{eqnarray}
The last two identities in Eq. (\ref{eq:id-1}) are results of following
identities 
\begin{eqnarray}
F^{\mu\nu}\tilde{F}_{\nu\lambda}p^{\lambda} & = & -\frac{1}{4}F^{\rho\sigma}\tilde{F}_{\rho\sigma}p^{\mu},\nonumber \\
\Omega^{\mu\nu}\tilde{\Omega}_{\nu\lambda}p^{\lambda} & = & -\frac{1}{4}\Omega^{\rho\sigma}\tilde{\Omega}_{\rho\sigma}p^{\mu}.\label{eq:id-2}
\end{eqnarray}
The proof of the first identity of Eq. (\ref{eq:id-2}) is as follows,
\begin{eqnarray}
F^{\mu\nu}\tilde{F}_{\nu\lambda}p^{\lambda} & = & -(p\cdot B)E^{\mu}-(E\cdot B)(u\cdot p)u^{\mu}-\epsilon^{\nu\mu\alpha\beta}\epsilon_{\nu\lambda\rho\sigma}u_{\alpha}B_{\beta}E^{\rho}u^{\sigma}p^{\lambda}\nonumber \\
 & = & -\frac{1}{4}F^{\rho\sigma}\tilde{F}_{\rho\sigma}p^{\mu},\label{eq:fft-1}
\end{eqnarray}
where we have used $u\cdot E=u\cdot B=0$, $4(E\cdot B)=F^{\rho\sigma}\tilde{F}_{\rho\sigma}$,
and 
\begin{eqnarray}
-\epsilon^{\nu\mu\alpha\beta}\epsilon_{\nu\lambda\rho\sigma}u_{\alpha}B_{\beta}E^{\rho}u^{\sigma}p^{\lambda} & = & (p\cdot B)E^{\mu}+(E\cdot B)(u\cdot p)u^{\mu}-(E\cdot B)p^{\mu}.
\end{eqnarray}
The proof of the second identity of (\ref{eq:id-2}) is similar.

\section{Derivation of the vorticity part of Eq. (\ref{eq:ccke-4})}

\label{sub:derive-ccke2}We rewrite the vorticity part in Eq. (\ref{eq:ccke-2}).
The vorticity part of $dx^{\mu}/d\tau$ is treated as 
\begin{eqnarray*}
I_{1}(\omega) & = & -s\frac{p_{0}}{p^{2}}\tilde{\Omega}^{\mu\lambda}p_{\lambda}\\
 & = & -s\frac{p_{0}}{p^{2}}(\omega^{\mu}u^{\nu}-\omega^{\nu}u^{\mu})p_{\nu}\\
 & = & -s\frac{p_{0}^{2}}{p^{2}}\omega^{\mu}+s\frac{p_{0}}{p^{2}}(p\cdot\omega)u^{\mu},\\
I_{2}(\omega) & = & \frac{s}{2}\omega^{\mu},
\end{eqnarray*}
where we have used $\varepsilon^{\alpha}=\Omega^{\alpha\beta}u_{\beta}=0$.
So we combine the two terms, 
\begin{equation}
I_{1}(\omega)+I_{2}(\omega)=s\left(\frac{1}{2}-\frac{p_{0}^{2}}{p^{2}}\right)\omega^{\mu}+s\frac{p_{0}}{p^{2}}(p\cdot\omega)u^{\mu}.
\end{equation}
We deal with each term in the vorticity part of $dp^{\mu}/d\tau$
as follows. The first term is vanishing, 
\begin{eqnarray}
I_{3}(\omega) & = & -\frac{s}{8}p^{\mu}\Omega_{\rho\lambda}\tilde{\Omega}^{\rho\lambda}\nonumber \\
 & = & -\frac{s}{16}p^{\mu}\epsilon^{\rho\lambda\alpha\beta}(\partial_{\rho}u_{\lambda})(\partial_{\alpha}u_{\beta})\nonumber \\
 & = & -\frac{s}{16}p^{\mu}\epsilon^{\rho\lambda\alpha\beta}\partial_{\rho}(u_{\lambda}\partial_{\alpha}u_{\beta})\nonumber \\
 & = & -\frac{s}{8}p^{\mu}\partial_{\rho}\omega^{\rho}=0.
\end{eqnarray}
The second term is rewritten as 
\begin{eqnarray}
I_{4}(\omega) & = & sQ\frac{1}{p^{2}}u^{\mu}\Omega_{\nu\lambda}p^{\lambda}\tilde{F}^{\nu\kappa}p_{\kappa}\nonumber \\
 & = & sQu^{\mu}(E\cdot\omega)-sQ\frac{p_{0}^{2}}{p^{2}}(E\cdot\omega)u^{\mu}-sQ\frac{1}{p^{2}}(p\cdot\omega)(p\cdot E)u^{\mu}.
\end{eqnarray}
The third term is rewritten as 
\begin{eqnarray}
I_{5}(\omega) & = & \frac{1}{2}sQF^{\mu\kappa}\tilde{\Omega}_{\kappa\lambda}u^{\lambda}=-\frac{1}{2}sQ(E\cdot\omega)u^{\mu}.
\end{eqnarray}
The fourth term is rewritten as 
\begin{eqnarray}
I_{6}(\omega) & = & -sQ\frac{p_{0}}{p^{2}}F^{\mu\nu}\tilde{\Omega}_{\nu\lambda}p^{\lambda}\nonumber \\
 & = & sQ\frac{p_{0}^{2}}{p^{2}}(E\cdot\omega)u^{\mu}+sQ\frac{1}{p^{2}}p_{0}(p\cdot\omega)E^{\mu},
\end{eqnarray}
where we have used $\varepsilon^{\alpha}=\Omega^{\alpha\beta}u_{\beta}=0$.
We combine these four terms as 
\begin{eqnarray}
\sum_{i=3}^{6}I_{i}(\omega) & = & \frac{1}{2}sQ(E\cdot\omega)u^{\mu}-sQ\frac{1}{p^{2}}(p\cdot\omega)(p\cdot E)u^{\mu}+sQ\frac{1}{p^{2}}p_{0}(p\cdot\omega)E^{\mu}.
\end{eqnarray}

\section{Derivation of Eq. (\ref{eq:y-vector})}

\label{sec:y-vect}In this appendix, we give a derivation of Eq. (\ref{eq:y-vector}).
Inserting Eq. (\ref{eq:x-mu}) into (\ref{eq:x-y-rel}) we obtain 

\begin{eqnarray}
0 & = & [sC_{1}(p,u)\omega^{\sigma}+sC_{2}(p,u)(p\cdot\omega)u^{\sigma}\nonumber \\
 &  & +sC_{3}(p,u)(p\cdot\omega)\bar{p}^{\sigma}]\partial_{\sigma}^{x}f_{s}+Y^{\sigma}\partial_{\sigma}^{p}f_{s}\nonumber \\
 & = & sC_{1}(p,u)\omega^{\sigma}[\partial_{\sigma}^{x}(u\cdot p)-\partial_{\sigma}^{x}\mu]\frac{df_{s}}{dp_{0}}\nonumber \\
 &  & +sC_{2}(p,u)(p\cdot\omega)u^{\sigma}[\partial_{\sigma}^{x}(u\cdot p)-\partial_{\sigma}^{x}\mu]\frac{df_{s}}{dp_{0}}\nonumber \\
 &  & +sC_{3}(p,u)(p\cdot\omega)\bar{p}^{\sigma}[\partial_{\sigma}^{x}(u\cdot p)-\partial_{\sigma}^{x}\mu]\frac{df_{s}}{dp_{0}}+(Y\cdot u)\frac{df_{s}}{dp_{0}}\nonumber \\
 & = & [sQC_{1}(p,u)(\omega\cdot E)+sQC_{3}(p,u)(p\cdot\omega)(p\cdot E)+(Y\cdot u)]\frac{df_{s}}{dp_{0}},
\end{eqnarray}
which leads to the following form of $Y^{\sigma}$ 
\begin{eqnarray}
Y^{\sigma} & = & -sQ[C_{1}(p,u)(\omega\cdot E)+C_{3}(p,u)(p\cdot\omega)(p\cdot E)]u^{\sigma}+\bar{Y}^{\sigma},\nonumber \\
Y^{\sigma}\partial_{\sigma}^{p}f_{s} & = & -sQ[C_{1}(p,u)(\omega\cdot E)+C_{3}(p,u)(p\cdot\omega)(p\cdot E)]u^{\sigma}\partial_{\sigma}^{p}f_{s},\label{eq:y-dfs}
\end{eqnarray}
where we have used $\bar{Y}^{\sigma}\equiv Y^{\sigma}-(Y\cdot u)u^{\sigma}$,
and we assume $\bar{Y}^{\sigma}$ has the form 
\begin{eqnarray}
\bar{Y}^{\sigma} & = & sQ\bar{p}^{\sigma}C_{4}(p,\omega),
\end{eqnarray}
with the function $C_{4}(p,\omega)$ being defined by 
\begin{equation}
C_{4}(p,\omega)=C_{40}(\omega\cdot E)\frac{1}{p_{0}}+C_{41}\frac{1}{p^{2}p_{0}}(p\cdot\omega)(p\cdot E),
\end{equation}
where $C_{40}$ and $C_{41}$ are two dimensionless coefficients to
be determined. We have used following identities derived from the
static and equilibrium conditions in (\ref{eq:static-equil}), 
\begin{eqnarray}
\partial_{\mu}^{p}f_{s} & = & u^{\mu}\frac{df_{s}}{dp_{0}},\nonumber \\
\omega^{\sigma}\partial_{\sigma}^{x}u_{\rho} & = & \omega^{\sigma}\Omega_{\sigma\rho}=\epsilon_{\sigma\rho\tau\mu}u^{\tau}\omega^{\mu}\omega^{\sigma}=0,\nonumber \\
u^{\sigma}\partial_{\sigma}^{x}f_{s} & = & u^{\sigma}[\partial_{\sigma}^{x}(u\cdot p)-\partial_{\sigma}^{x}\mu]\frac{df_{s}}{dp_{0}}=0,\nonumber \\
\bar{p}^{\sigma}p^{\rho}\partial_{\sigma}^{x}u_{\rho} & = & \bar{p}^{\sigma}\bar{p}^{\rho}\partial_{\sigma}^{x}u_{\rho}=\frac{1}{2}\bar{p}^{\sigma}\bar{p}^{\rho}\Omega_{\sigma\rho}=0.
\end{eqnarray}

\section{Derivation of formula in Section (\ref{sec:constraint}) }

\label{sec:der-curr}In this Appendix, we give detailed derivations
of currents and energy momentum tensor in Section (\ref{sec:constraint}).

\subsection{Currents}

For the electromagnetic field part of $dx^{\mu}/d\tau$ in Eq. (\ref{eq:new-ccke}),
we evaluate the current as 
\begin{eqnarray}
j_{s}^{\mu}(\mathrm{EM}) & = & -sQ\int d^{4}p\delta(p^{2})\frac{1}{p^{2}}\tilde{F}^{\mu\lambda}p_{\lambda}f_{s}\nonumber \\
 & = & \frac{1}{2}sQ\int d^{4}p\frac{d\delta(p^{2})}{dp_{0}}\frac{1}{p_{0}}(B^{\mu}p_{0}-(p\cdot B)u^{\mu}+\epsilon^{\mu\lambda\rho\sigma}p_{\lambda}E_{\rho}u_{\sigma})f_{s}\nonumber \\
 & = & -\frac{1}{2}sQB^{\mu}\int d^{4}p\delta(p^{2})\frac{df_{s}}{dp_{0}}\nonumber \\
 & = & -\frac{1}{2}sQB^{\mu}\int\frac{d^{3}p}{(2\pi)^{3}}\frac{1}{E_{p}}\frac{d}{dE_{p}}\left[f_{\mathrm{FD}}(E_{p}-\mu_{s})-f_{\mathrm{FD}}(E_{p}+\mu_{s})\right]\nonumber \\
 & = & B^{\mu}\frac{sQ}{4\pi^{2}}\int_{0}^{\infty}dE_{p}\left[f_{\mathrm{FD}}(E_{p}-\mu_{s})-f_{\mathrm{FD}}(E_{p}+\mu_{s})\right]\nonumber \\
 & \equiv & \xi_{B}^{s}B^{\mu},\label{eq:j-em}
\end{eqnarray}
where we have dropped the complete derivative in $p_{0}$ and those
integrals whose integrands are odd in $\bar{p}$. This gives the derivation
of Eq. (\ref{eq:js-em}). 

Now we derive Eq. (\ref{eq:curr-vorticity-part}) for the currents
induced by the vorticity. We put the vorticity part of $dx^{\mu}/d\tau$
in Eq. (\ref{eq:new-ccke}) into Eq. (\ref{eq:curr-vort}) and obtain
\begin{eqnarray}
j_{s}^{\mu}(\omega) & = & s\left(\frac{1}{2}+C_{10}\right)\omega^{\mu}\int d^{4}p\delta(p^{2})f_{s}-\frac{1}{2}s(C_{11}-1)\omega^{\mu}\int d^{4}pp_{0}\frac{d\delta(p^{2})}{dp_{0}}f_{s}\nonumber \\
 &  & -\frac{1}{2}sC_{30}\omega_{\lambda}\int d^{4}p\bar{p}^{\lambda}\bar{p}^{\mu}\frac{d\delta(p^{2})}{dp_{0}}\frac{1}{p_{0}}f_{s}\nonumber \\
 & = & s\left(C_{10}-\frac{1}{2}C_{11}+1\right)\omega^{\mu}\int d^{4}p\delta(p^{2})f_{s}+\frac{1}{2}sC_{30}\omega^{\mu}\int d^{4}p\delta(p^{2})f_{s}\nonumber \\
 & = & \left(C_{10}-\frac{1}{2}C_{11}+\frac{1}{2}C_{30}+1\right)\omega^{\mu}\int d^{4}p\delta(p^{2})sf_{s}\nonumber \\
 & = & \left(C_{10}-\frac{1}{2}C_{11}+\frac{1}{2}C_{30}+1\right)\xi_{s}\omega^{\mu},\label{eq:der-j-omega}
\end{eqnarray}
where we have defined (no summation over $s$ is implied) 
\begin{eqnarray}
\xi_{s} & \equiv & \int d^{4}p\delta(p^{2})sf_{s}\nonumber \\
 & = & \frac{1}{2\pi^{2}}s\int_{0}^{\infty}dE_{p}E_{p}\left[f_{\mathrm{FD}}(E_{p}-\mu_{s})+f_{\mathrm{FD}}(E_{p}+\mu_{s})\right].
\end{eqnarray}
In the first equality of Eq. (\ref{eq:der-j-omega}) we have used
the fact that $f_{s}$ is given by Eq. (\ref{eq:dist}) and depends
on momentum through $p_{0}=p\cdot u$, so the integral proportional
to $u^{\sigma}$ is vanishing since the integrand is odd in spatial
momentum due to $p\cdot\omega=\bar{p}\cdot\omega$. We have dropped
complete integrals in $p_{0}$ in Eq. (\ref{eq:der-j-omega}). We
have also used following integrals, 
\begin{eqnarray}
\int d^{4}p\delta(p^{2})f_{s} & = & \int\frac{d^{3}p}{(2\pi)^{3}}\frac{1}{E_{p}}\left[f_{\mathrm{FD}}(E_{p}-\mu_{s})+f_{\mathrm{FD}}(E_{p}+\mu_{s})\right],\nonumber \\
\int d^{4}pp_{0}\frac{d\delta(p^{2})}{dp_{0}}f_{s} & = & -\int d^{4}p\delta(p^{2})f_{s}-\int d^{4}p\delta(p^{2})p_{0}\frac{df_{s}}{dp_{0}}\nonumber \\
 & = & \int d^{4}p\delta(p^{2})f_{s},\nonumber \\
\int d^{4}p\bar{p}^{\lambda}\bar{p}^{\mu}\frac{d\delta(p^{2})}{dp_{0}}\frac{1}{p_{0}}f_{s} & = & \int d^{4}p\bar{p}^{\lambda}\bar{p}^{\mu}\delta(p^{2})\frac{1}{p_{0}^{2}}f_{s}-\int d^{4}p\delta(p^{2})\bar{p}^{\lambda}\bar{p}^{\mu}\frac{1}{p_{0}}\frac{df_{s}}{dp_{0}}\nonumber \\
 & = & -\Delta^{\lambda\mu}\int d^{4}p\delta(p^{2})f_{s}.
\end{eqnarray}
Here we have carried out the momentum integrals in the co-moving frame
of the fluid with $u^{\mu}=(1,0)$ and $\bar{p}^{\mu}=(0,\mathbf{p})$.
We have used the assumption that $f_{s}$ depends on $\mathbf{p}$
through $u\cdot p$ so it is isotropic in spatial momentum, $\bar{p}^{\sigma}\bar{p}^{\rho}\rightarrow-\frac{1}{3}|\mathbf{p}|^{2}\Delta^{\sigma\rho}$.
We have also dropped complete integrals in $p_{0}$ and $E_{p}$.

\subsection{Energy-momentum tensor}

We can derive the energy-momentum tensor in Eq. (\ref{eq:emt}) in
a similar way to how we derive the currents in the last subsection. 

We evaluate the electromagnetic field part $T^{\rho\sigma}(\mathrm{EM})$
as 
\begin{eqnarray}
T^{\rho\sigma}(\mathrm{EM}) & = & -\frac{1}{2}sQ\int d^{4}p\delta(p^{2})\frac{1}{p^{2}}\left(p^{\rho}p_{\lambda}\tilde{F}^{\sigma\lambda}f_{s}+p^{\sigma}p_{\lambda}\tilde{F}^{\rho\lambda}f_{s}\right)\nonumber \\
 & = & \frac{1}{4}sQ\int d^{4}p\frac{d\delta(p^{2})}{dp_{0}}\frac{1}{p_{0}}f_{s}\left(p^{\rho}p_{\lambda}\tilde{F}^{\sigma\lambda}+p^{\sigma}p_{\lambda}\tilde{F}^{\rho\lambda}\right)\nonumber \\
 & = & -\frac{1}{3}sQu^{(\rho}B^{\sigma)}\left[\frac{1}{2}\int d^{4}p\delta(p^{2})f_{s}+\int d^{4}p\delta(p^{2})\frac{|\mathbf{p}|^{2}}{p_{0}}\frac{df_{s}}{dp_{0}}\right]\nonumber \\
 & = & \frac{1}{2}sQu^{(\rho}B^{\sigma)}\frac{1}{2\pi^{2}}\int_{0}^{\infty}dE_{p}E_{p}\left[f_{\mathrm{FD}}(E_{p}-\mu_{s})+f_{\mathrm{FD}}(E_{p}+\mu_{s})\right]\nonumber \\
 & = & \frac{1}{2}Q\xi u^{(\rho}B^{\sigma)},\label{eq:t-em}
\end{eqnarray}
which reproduces the previous result \cite{Gao:2012ix}. Here the
summation over $s$ was implied and we have used 
\begin{eqnarray}
p^{\rho}p_{\lambda} & = & p_{0}^{2}u^{\rho}u_{\lambda}+\bar{p}^{\rho}\bar{p}_{\lambda}+\cdots\nonumber \\
 & \rightarrow & p_{0}^{2}u^{\rho}u_{\lambda}-|\mathbf{p}|^{2}\frac{1}{3}\Delta_{\lambda}^{\rho},\label{eq:p-rho-p-l}
\end{eqnarray}
where we have replaced $\bar{p}^{\rho}\bar{p}_{\lambda}\rightarrow-|\mathbf{p}|^{2}\frac{1}{3}\Delta_{\lambda}^{\rho}$
which is true in 3D momentum integrals for isotropic momentum distributions.
Note that we have not shown terms linear in $\bar{p}$ in Eq. (\ref{eq:p-rho-p-l})
which give vanishing integrals. In Eq. (\ref{eq:t-em}) we also used
the integral 
\begin{eqnarray}
\int d^{4}p\delta(p^{2})f_{s} & = & \int\frac{d^{3}p}{(2\pi)^{3}}\frac{1}{E_{p}}\left[f_{\mathrm{FD}}(E_{p}-\mu_{s})+f_{\mathrm{FD}}(E_{p}+\mu_{s})\right],\nonumber \\
\int d^{4}p\delta(p^{2})\frac{|\mathbf{p}|^{2}}{p_{0}}\frac{df_{s}}{dp_{0}} & = & \int\frac{d^{3}p}{(2\pi)^{3}}\frac{d}{dE_{p}}\left[f_{\mathrm{FD}}(E_{p}-\mu_{s})+f_{\mathrm{FD}}(E_{p}+\mu_{s})\right]\nonumber \\
 & = & -2\int\frac{d^{3}p}{(2\pi)^{3}}\frac{1}{E_{p}}\left[f_{\mathrm{FD}}(E_{p}-\mu_{s})+f_{\mathrm{FD}}(E_{p}+\mu_{s})\right].
\end{eqnarray}
Here we have dropped the complete integral in $E_{p}$ which is vanishing. 

For the vorticity part, we insert the last three terms of $dx^{\mu}/d\tau$
in Eq. (\ref{eq:new-ccke}) into Eq. (\ref{eq:stress-t}) and obtain
\begin{eqnarray}
T^{\rho\sigma}(\omega) & = & \frac{1}{2}\left(\frac{1}{2}+C_{10}\right)u^{(\rho}\omega^{\sigma)}\int d^{4}pp_{0}\delta(p^{2})sf_{s}-\frac{1}{4}(C_{11}-1)\int d^{4}p\frac{d\delta(p^{2})}{dp_{0}}p_{0}p^{(\rho}\omega^{\sigma)}sf_{s}\nonumber \\
 &  & -\frac{1}{4}(C_{20}+1)\int d^{4}p\frac{d\delta(p^{2})}{dp_{0}}p^{(\rho}u^{\sigma)}(p\cdot\omega)sf_{s}+\frac{1}{2}C_{21}\int d^{4}p\delta(p^{2})\frac{1}{p_{0}}(p\cdot\omega)p^{(\rho}u^{\sigma)}sf_{s}\nonumber \\
 &  & -\frac{1}{4}C_{30}\int d^{4}p\frac{d\delta(p^{2})}{dp_{0}}\frac{1}{p_{0}}(\bar{p}\cdot\omega)p^{(\rho}\bar{p}^{\sigma)}sf_{s}\nonumber \\
 & = & \frac{1}{2}\left(\frac{1}{2}+C_{10}\right)u^{(\rho}\omega^{\sigma)}n_{5}-\frac{1}{4}(C_{11}-1)u^{(\rho}\omega^{\sigma)}n_{5}\nonumber \\
 &  & +\frac{1}{4}(C_{20}+1)u^{(\rho}\omega^{\sigma)}n_{5}-\frac{1}{6}C_{21}u^{(\rho}\omega^{\sigma)}n_{5}+\frac{1}{4}C_{30}u^{(\rho}\omega^{\sigma)}n_{5}\nonumber \\
 & = & n_{5}u^{(\rho}\omega^{\sigma)}\left(\frac{1}{2}C_{10}-\frac{1}{4}C_{11}+\frac{1}{4}C_{30}+\frac{1}{4}C_{20}-\frac{1}{6}C_{21}+\frac{3}{4}\right),\label{eq:stress-vorticity}
\end{eqnarray}
where the summation over $s$ was implied and we have used following
integrals 
\begin{eqnarray}
\int d^{4}pp_{0}\delta(p^{2})sf_{s} & = & \int\frac{d^{3}p}{(2\pi)^{3}}s\left[f_{\mathrm{FD}}(E_{p}-\mu_{s})-f_{\mathrm{FD}}(E_{p}+\mu_{s})\right]=n_{5},\nonumber \\
\int d^{4}p\frac{d\delta(p^{2})}{dp_{0}}p_{0}^{2}sf_{s} & = & -2n_{5}-\int d^{4}p\delta(p^{2})p_{0}^{2}\frac{d}{dp_{0}}sf_{s}=n_{5},\nonumber \\
\int d^{4}p\frac{d\delta(p^{2})}{dp_{0}}p^{(\rho}u^{\sigma)}(p\cdot\omega)sf_{s} & = & -\int d^{4}p\bar{p}^{(\rho}u^{\sigma)}(\bar{p}\cdot\omega)\delta(p^{2})\frac{d(sf_{s})}{dp_{0}}\nonumber \\
 & = & \frac{1}{3}\omega_{\mu}u^{(\sigma}\Delta^{\rho)\mu}\int d^{4}pE_{p}^{2}\delta(p^{2})\frac{d(sf_{s})}{dp_{0}}\nonumber \\
 & = & -u^{(\sigma}\omega^{\rho)}n_{5},\nonumber \\
\int d^{4}p\delta(p^{2})\frac{1}{p_{0}}(\bar{p}\cdot\omega)\bar{p}^{(\rho}u^{\sigma)}sf_{s} & = & -\frac{1}{3}\omega_{\lambda}\Delta^{\lambda(\rho}u^{\sigma)}\int d^{4}p\delta(p^{2})E_{p}^{2}\frac{1}{p_{0}}sf_{s}\nonumber \\
 & = & -\frac{1}{3}n_{5}u^{(\rho}\omega^{\sigma)},\nonumber \\
\int d^{4}p\frac{d\delta(p^{2})}{dp_{0}}\frac{1}{p_{0}}(\bar{p}\cdot\omega)p^{(\rho}\bar{p}^{\sigma)}sf_{s} & = & \frac{1}{3}\omega_{\lambda}u^{(\rho}\Delta^{\lambda\sigma)}\int d^{4}p\delta(p^{2})E_{p}^{2}\frac{d(sf_{s})}{dp_{0}}\nonumber \\
 & = & -n_{5}u^{(\rho}\omega^{\sigma)}.
\end{eqnarray}
For comparison, we can obtain $T^{\rho\sigma}(\omega)$ from $\mathscr{J}_{(1)s}^{\rho}(x,p)$
in Eq. (\ref{eq:1st-solution}) by the definition in Ref. \cite{Gao:2012ix},
\begin{eqnarray}
T^{\rho\sigma}(\omega) & = & \frac{1}{2}\int d^{4}p[p^{\rho}\mathscr{J}_{(1)s}^{\sigma}+p^{\sigma}\mathscr{J}_{(1)s}^{\rho}]\nonumber \\
 & \rightarrow & -\frac{s}{4}\int d^{4}p\left(p^{\rho}p_{\beta}\tilde{\Omega}^{\sigma\beta}+p^{\sigma}p_{\beta}\tilde{\Omega}^{\rho\beta}\right)\frac{df_{s}}{dp_{0}}\delta(p^{2})\nonumber \\
 & = & -\frac{s}{4}\int d^{4}p\left[(p_{0}^{2}u^{\sigma}u_{\beta}-\frac{1}{3}E_{p}^{2}\Delta_{\beta}^{\sigma})\tilde{\Omega}^{\rho\beta}+(p_{0}^{2}u^{\rho}u_{\beta}-\frac{1}{3}E_{p}^{2}\Delta_{\beta}^{\rho})\tilde{\Omega}^{\sigma\beta}\right]\frac{df_{s}}{dp_{0}}\delta(p^{2})\nonumber \\
 & = & -\frac{s}{3}\int d^{4}pE_{p}^{2}\left(u^{\rho}u_{\beta}\tilde{\Omega}^{\sigma\beta}+u^{\sigma}u_{\beta}\tilde{\Omega}^{\rho\beta}\right)\frac{df_{s}}{dp_{0}}\delta(p^{2})\nonumber \\
 & = & u^{(\sigma}\omega^{\rho)}\int\frac{d^{3}p}{(2\pi)^{3}}s\left[f_{\mathrm{FD}}(E_{p}-\mu_{s})-f_{\mathrm{FD}}(E_{p}+\mu_{s})\right]\nonumber \\
 & = & n_{5}u^{(\sigma}\omega^{\rho)},\label{eq:der-t-vort}
\end{eqnarray}
where we also used Eq. (\ref{eq:p-rho-p-l}) and the summation over
$s=\pm1$ is implied.

\section{Evaluation of each term in Eq. (\ref{eq:ix-ip})}

\label{sec:derivation-ccke}In this appendix we evaluate each term
in Eq. (\ref{eq:ix-ip}) with $dx^{\mu}/d\tau$ and $dp^{\mu}/d\tau$
given by Eq. (\ref{eq:new-ccke}). We work in the co-moving frame with
$u^{\mu}=(1,0)$.

\subsection{Evaluation of $I_{x0}$}

The term $I_{x0}$ consists of three parts, $I_{x0}(0)$ of the zeroth
order contribution, $I_{x0}(\mathrm{EM})$ of the first order in electromagnetic
field, and $I_{x0}(\omega)$ of the first order in vorticity. The
detailed derivation is as follows, 
\begin{eqnarray}
I_{x0}(0) & = & \int dp_{0}\delta(p^{2})p_{0}\partial_{0}^{x}f_{s}\nonumber \\
 & = & \frac{1}{(2\pi)^{3}}\partial_{0}^{x}\left[f_{\mathrm{FD}}(E_{p}-\mu_{s})-f_{\mathrm{FD}}(E_{p}+\mu_{s})\right],\label{eq:ix0}\\
I_{x0}(\mathrm{EM}) & = & sQ\int dp_{0}\frac{1}{p^{2}}\delta(p^{2})(p\cdot B)\partial_{0}^{x}f_{s}\nonumber \\
 & = & \frac{1}{2}sQ(\mathbf{p}\cdot\mathbf{B})\int dp_{0}\frac{d\delta(p^{2})}{dp_{0}}\frac{1}{p_{0}}\partial_{0}^{x}f_{s}\nonumber \\
 & = & \frac{1}{(2\pi)^{3}}\left\{ \frac{1}{2}sQ(\mathbf{p}\cdot\mathbf{B})\frac{1}{E_{p}^{3}}\partial_{0}^{x}\left[f_{\mathrm{FD}}(E_{p}-\mu_{s})+f_{\mathrm{FD}}(E_{p}+\mu_{s})\right]\right.\nonumber \\
 &  & \left.-\frac{1}{2}sQ(\mathbf{p}\cdot\mathbf{B})\frac{1}{E_{p}^{2}}\frac{d}{dE_{p}}\partial_{0}^{x}\left[f_{\mathrm{FD}}(E_{p}-\mu_{s})+f_{\mathrm{FD}}(E_{p}+\mu_{s})\right]\right\} ,\label{eq:ix0-em}\\
I_{x0}(\omega) & = & s\int dp_{0}\delta(p^{2})\left[(C_{20}+1)\frac{p_{0}}{p^{2}}+C_{21}\frac{1}{p_{0}}\right](p\cdot\omega)\partial_{0}^{x}f_{s}\nonumber \\
 & = & \frac{1}{2}s(C_{20}+1)(\mathbf{p}\cdot\boldsymbol{\omega})\int dp_{0}\frac{d\delta(p^{2})}{dp_{0}}\partial_{0}^{x}f_{s}-sC_{21}(\mathbf{p}\cdot\boldsymbol{\omega})\int dp_{0}\delta(p^{2})\frac{1}{p_{0}}\partial_{0}^{x}f_{s}\nonumber \\
 & \rightarrow & \frac{1}{(2\pi)^{3}}s(C_{20}-C_{21}+1)(\mathbf{p}\cdot\boldsymbol{\omega})\frac{1}{E_{p}^{2}}\partial_{0}^{x}\left[f_{\mathrm{FD}}(E_{p}-\mu_{s})-f_{\mathrm{FD}}(E_{p}+\mu_{s})\right],\label{eq:ix0-vort}
\end{eqnarray}
where the involved integrals are evaluated as 
\begin{eqnarray}
\int dp_{0}\frac{d\delta(p^{2})}{dp_{0}}\frac{1}{p_{0}}\partial_{0}^{x}f_{s} & = & \int dp_{0}\delta(p^{2})\frac{1}{p_{0}^{2}}\partial_{0}^{x}f_{s}-\int dp_{0}\delta(p^{2})\frac{1}{p_{0}}\frac{d}{dp_{0}}\partial_{0}^{x}f_{s}\nonumber \\
 & = & \frac{1}{(2\pi)^{3}}\left\{ \frac{1}{E_{p}^{3}}\partial_{0}^{x}\left[f_{\mathrm{FD}}(E_{p}-\mu_{s})+f_{\mathrm{FD}}(E_{p}+\mu_{s})\right]\right.\nonumber \\
 &  & \left.-\frac{1}{E_{p}^{2}}\frac{d}{dE_{p}}\partial_{0}^{x}\left[f_{\mathrm{FD}}(E_{p}-\mu_{s})+f_{\mathrm{FD}}(E_{p}+\mu_{s})\right]\right\} ,\label{eq:int-p-b}\\
\int dp_{0}\frac{d\delta(p^{2})}{dp_{0}}\partial_{0}^{x}f_{s} & = & -\int dp_{0}\delta(p^{2})\frac{d}{dp_{0}}\partial_{0}^{x}f_{s}\nonumber \\
 & = & -\frac{1}{(2\pi)^{3}}\frac{1}{E_{p}}\frac{d}{dE_{p}}\partial_{0}^{x}\left[f_{\mathrm{FD}}(E_{p}-\mu_{s})-f_{\mathrm{FD}}(E_{p}+\mu_{s})\right],\;(E_{p}\:\mathrm{in\: integrand})\nonumber \\
 & \rightarrow & \frac{1}{(2\pi)^{3}}\frac{2}{E_{p}^{2}}\partial_{0}^{x}\left[f_{\mathrm{FD}}(E_{p}-\mu_{s})-f_{\mathrm{FD}}(E_{p}+\mu_{s})\right],\label{eq:int-vort}\\
\int dp_{0}\delta(p^{2})\frac{1}{p_{0}}\partial_{0}^{x}f_{s} & = & \frac{1}{(2\pi)^{3}}\frac{1}{E_{p}^{2}}\partial_{0}^{x}\left[f_{\mathrm{FD}}(E_{p}-\mu_{s})-f_{\mathrm{FD}}(E_{p}+\mu_{s})\right].
\end{eqnarray}
In the last line of the integral (\ref{eq:int-vort}), we have implied
that there will be a $\int d^{3}p$ integral so we have 
\begin{eqnarray}
 &  & \int d^{3}p(\mathbf{p}\cdot\boldsymbol{\omega})\int dp_{0}\frac{d\delta(p^{2})}{dp_{0}}\partial_{0}^{x}f_{s}\nonumber \\
 & \rightarrow & -\int d\Omega_{p}(\hat{\mathbf{p}}\cdot\boldsymbol{\omega})\int_{0}^{\infty}dE_{p}E_{p}^{2}E_{p}\frac{1}{E_{p}}\frac{d}{dE_{p}}\partial_{0}^{x}\left[f_{\mathrm{FD}}(E_{p}-\mu_{s})-f_{\mathrm{FD}}(E_{p}+\mu_{s})\right]\nonumber \\
 & = & -\int d\Omega_{p}(\hat{\mathbf{p}}\cdot\boldsymbol{\omega})\int_{0}^{\infty}dE_{p}\frac{d}{dE_{p}}\left\{ E_{p}^{2}\partial_{0}^{x}\left[f_{\mathrm{FD}}(E_{p}-\mu_{s})-f_{\mathrm{FD}}(E_{p}+\mu_{s})\right]\right\} \nonumber \\
 &  & +2\int d\Omega_{p}(\hat{\mathbf{p}}\cdot\boldsymbol{\omega})\int_{0}^{\infty}dE_{p}E_{p}\partial_{0}^{x}\left[f_{\mathrm{FD}}(E_{p}-\mu_{s})-f_{\mathrm{FD}}(E_{p}+\mu_{s})\right]\nonumber \\
 & \sim & \int d^{3}p(\mathbf{p}\cdot\boldsymbol{\omega})\frac{2}{E_{p}^{2}}\partial_{0}^{x}\left[f_{\mathrm{FD}}(E_{p}-\mu_{s})-f_{\mathrm{FD}}(E_{p}+\mu_{s})\right],\label{eq:complete-int}
\end{eqnarray}
where the complete integral term in $E_{p}$ is vanishing at two limits
$E_{p}=0,\infty$. The above integral can also be treated in a slightly
different way,
\begin{eqnarray}
 &  & \int d^{3}p(\mathbf{p}\cdot\boldsymbol{\omega})\int dp_{0}\frac{d\delta(p^{2})}{dp_{0}}\partial_{0}^{x}f_{s}\nonumber \\
 & \rightarrow & -\int d^{3}p(\mathbf{p}\cdot\boldsymbol{\omega})\frac{1}{E_{p}}\frac{d}{dE_{p}}\partial_{0}^{x}\left[f_{\mathrm{FD}}(E_{p}-\mu_{s})-f_{\mathrm{FD}}(E_{p}+\mu_{s})\right]\nonumber \\
 & = & -\int d^{3}p\frac{1}{E_{p}^{2}}(\mathbf{p}\cdot\boldsymbol{\omega})\mathbf{p}\cdot\nabla_{p}\partial_{0}^{x}\left[f_{\mathrm{FD}}(E_{p}-\mu_{s})-f_{\mathrm{FD}}(E_{p}+\mu_{s})\right]\nonumber \\
 & = & -\int d^{3}p\nabla_{p}\cdot\left\{ \frac{1}{E_{p}^{2}}(\mathbf{p}\cdot\boldsymbol{\omega})\mathbf{p}\partial_{0}^{x}\left[f_{\mathrm{FD}}(E_{p}-\mu_{s})-f_{\mathrm{FD}}(E_{p}+\mu_{s})\right]\right\} \nonumber \\
 &  & +\int d^{3}p\nabla_{p}\cdot\left[\frac{1}{E_{p}^{2}}(\mathbf{p}\cdot\boldsymbol{\omega})\mathbf{p}\right]\partial_{0}^{x}\left[f_{\mathrm{FD}}(E_{p}-\mu_{s})-f_{\mathrm{FD}}(E_{p}+\mu_{s})\right]\nonumber \\
 & = & \int d^{3}p(\mathbf{p}\cdot\boldsymbol{\omega})\frac{2}{E_{p}^{2}}\partial_{0}^{x}\left[f_{\mathrm{FD}}(E_{p}-\mu_{s})-f_{\mathrm{FD}}(E_{p}+\mu_{s})\right],\label{eq:complete-int-1}
\end{eqnarray}
where we have used 
\begin{eqnarray}
\frac{d}{dE_{p}} & \rightarrow & \frac{1}{E_{p}}\mathbf{p}\cdot\nabla_{p},\nonumber \\
\nabla_{p}\cdot\left[\frac{1}{E_{p}^{2}}(\mathbf{p}\cdot\boldsymbol{\omega})\mathbf{p}\right] & = & (\mathbf{p}\cdot\boldsymbol{\omega})\nabla_{p}\cdot\left(\frac{\mathbf{p}}{E_{p}^{2}}\right)+\frac{1}{E_{p}^{2}}\mathbf{p}\cdot\nabla_{p}(\mathbf{p}\cdot\boldsymbol{\omega})\nonumber \\
 & = & \frac{2}{E_{p}^{2}}(\mathbf{p}\cdot\boldsymbol{\omega}),
\end{eqnarray}
and the total divergence term is vanishing.

\subsection{Evaluation of $I_{x}$}

Now we work on $I_{x}=I_{x}(0)+I_{x}(\mathrm{EM})+I_{x}(\omega)$,
where $I_{x}(0)$ is the the zeroth order contribution, and $I_{x}(\mathrm{EM})$
and $I_{x}(\omega)$ are the first order contribution from electromagnetic
field and vorticity respectively. We evaluate these terms as 
\begin{eqnarray}
I_{x}(0) & = & \int dp_{0}\delta(p^{2})p^{i}\partial_{i}^{x}f_{s}\nonumber \\
 & = & \frac{\mathbf{p}_{i}}{E_{p}}\frac{1}{(2\pi)^{3}}\partial_{i}^{x}\left[f_{\mathrm{FD}}(E_{p}-\mu_{s})+f_{\mathrm{FD}}(E_{p}+\mu_{s})\right],\label{eq:ix}\\
I_{x}(\mathrm{EM}) & = & -sQB^{i}\int dp_{0}p_{0}\frac{1}{p^{2}}\delta(p^{2})\partial_{i}f_{s}+sQ\epsilon^{0ijk}\int dp_{0}\frac{1}{p^{2}}\delta(p^{2})p_{j}E_{k}\partial_{i}f_{s}\nonumber \\
 & = & \frac{1}{2}sQ\mathbf{B}_{i}\int dp_{0}\frac{d\delta(p^{2})}{dp_{0}}\partial_{i}f_{s}-\frac{1}{2}sQ(\mathbf{p}\times\mathbf{E})_{i}\int dp_{0}\frac{d\delta(p^{2})}{dp_{0}}\frac{1}{p_{0}}\partial_{i}f_{s}\nonumber \\
 & = & \frac{1}{(2\pi)^{3}}\frac{1}{2}sQ\mathbf{B}_{i}\frac{1}{E_{p}^{2}}\partial_{i}^{x}\left[f_{\mathrm{FD}}(E_{p}-\mu_{s})-f_{\mathrm{FD}}(E_{p}+\mu_{s})\right]\nonumber \\
 &  & -\frac{1}{(2\pi)^{3}}\frac{1}{2}sQ(\mathbf{p}\times\mathbf{E})_{i}\frac{1}{E_{p}^{3}}\partial_{i}^{x}\left[f_{\mathrm{FD}}(E_{p}-\mu_{s})+f_{\mathrm{FD}}(E_{p}+\mu_{s})\right]\nonumber \\
 &  & +\frac{1}{(2\pi)^{3}}\frac{1}{2}sQ(\mathbf{p}\times\mathbf{E})_{i}\frac{1}{E_{p}^{2}}\frac{d}{dE_{p}}\partial_{i}^{x}\left[f_{\mathrm{FD}}(E_{p}-\mu_{s})+f_{\mathrm{FD}}(E_{p}+\mu_{s})\right],\label{eq:ix-em}\\
I_{x}(\omega) & = & s\int dp_{0}\delta(p^{2})\left[\frac{1}{2}+C_{10}+(C_{11}-1)\frac{p_{0}^{2}}{p^{2}}\right]\omega^{i}\partial_{i}f_{s}\nonumber \\
 &  & +sC_{30}\int dp_{0}\delta(p^{2})\frac{1}{p^{2}}(p\cdot\omega)\bar{p}^{i}\partial_{i}f_{s}\nonumber \\
 & = & \frac{1}{(2\pi)^{3}}s\left(\frac{1}{2}+C_{10}\right)\frac{1}{E_{p}}\omega^{i}\partial_{i}^{x}\left[f_{\mathrm{FD}}(E_{p}-\mu_{s})+f_{\mathrm{FD}}(E_{p}+\mu_{s})\right]\nonumber \\
 &  & -\frac{1}{2}s(C_{11}-1)\omega^{i}\int dp_{0}\delta(p^{2})\frac{d\delta(p^{2})}{dp_{0}}p_{0}\partial_{i}f_{s}\nonumber \\
 &  & -\frac{1}{2}sC_{30}\bar{p}^{j}\bar{p}^{i}\omega_{j}\int dp_{0}\delta(p^{2})\frac{d\delta(p^{2})}{dp_{0}}\frac{1}{p_{0}}\partial_{i}f_{s}\nonumber \\
 & = & \frac{1}{(2\pi)^{3}}s\left(\frac{1}{2}+C_{10}\right)\frac{1}{E_{p}}\omega^{i}\partial_{i}^{x}\left[f_{\mathrm{FD}}(E_{p}-\mu_{s})+f_{\mathrm{FD}}(E_{p}+\mu_{s})\right]\nonumber \\
 &  & -\frac{1}{(2\pi)^{3}}s(C_{11}-1)\frac{1}{2E_{p}}\omega^{i}\partial_{i}^{x}\left[f_{\mathrm{FD}}(E_{p}-\mu_{s})+f_{\mathrm{FD}}(E_{p}+\mu_{s})\right]\nonumber \\
 &  & -\frac{1}{(2\pi)^{3}}sC_{30}\bar{p}^{j}\bar{p}^{i}\omega_{j}\frac{3}{2E_{p}^{3}}\partial_{i}^{x}\left[f_{\mathrm{FD}}(E_{p}-\mu_{s})+f_{\mathrm{FD}}(E_{p}+\mu_{s})\right]\nonumber \\
 & = & \frac{1}{(2\pi)^{3}}s\left(1+C_{10}-\frac{1}{2}C_{11}\right)\frac{1}{E_{p}}\boldsymbol{\omega}^{i}\partial_{i}^{x}\left[f_{\mathrm{FD}}(E_{p}-\mu_{s})+f_{\mathrm{FD}}(E_{p}+\mu_{s})\right]\nonumber \\
 &  & +\frac{1}{(2\pi)^{3}}sC_{30}(\mathbf{p}\cdot\boldsymbol{\mathbf{\omega}})\mathbf{p}_{i}\frac{3}{2E_{p}^{3}}\partial_{i}^{x}\left[f_{\mathrm{FD}}(E_{p}-\mu_{s})+f_{\mathrm{FD}}(E_{p}+\mu_{s})\right],\label{eq:ix-vort}
\end{eqnarray}
where we have used following integrals 
\begin{eqnarray}
\int dp_{0}\frac{d\delta(p^{2})}{dp_{0}}\partial_{i}f_{s} & = & -\int dp_{0}\delta(p^{2})\frac{d}{dp_{0}}\partial_{i}f_{s}\nonumber \\
 & = & -\frac{1}{E_{p}}\frac{d}{dE_{p}}\partial_{i}^{x}\left[f_{\mathrm{FD}}(E_{p}-\mu_{s})-f_{\mathrm{FD}}(E_{p}+\mu_{s})\right],\;(E_{p}^{0}\:\mathrm{in\: integrand})\nonumber \\
 & \rightarrow & \frac{1}{E_{p}^{2}}\frac{1}{(2\pi)^{3}}\partial_{i}^{x}\left[f_{\mathrm{FD}}(E_{p}-\mu_{s})-f_{\mathrm{FD}}(E_{p}+\mu_{s})\right],\label{eq:em-int-ix-b}\\
\int dp_{0}\frac{d\delta(p^{2})}{dp_{0}}\frac{1}{p_{0}}\partial_{i}f_{s} & = & \int dp_{0}\delta(p^{2})\frac{1}{p_{0}^{2}}\partial_{i}f_{s}-\int dp_{0}\delta(p^{2})\frac{1}{p_{0}}\frac{d}{dp_{0}}\partial_{i}f_{s}\nonumber \\
 & = & \frac{1}{(2\pi)^{3}}\frac{1}{E_{p}^{3}}\partial_{i}^{x}\left[f_{\mathrm{FD}}(E_{p}-\mu_{s})+f_{\mathrm{FD}}(E_{p}+\mu_{s})\right]\nonumber \\
 &  & -\frac{1}{(2\pi)^{3}}\frac{1}{E_{p}^{2}}\frac{d}{dE_{p}}\partial_{i}^{x}\left[f_{\mathrm{FD}}(E_{p}-\mu_{s})+f_{\mathrm{FD}}(E_{p}+\mu_{s})\right],\label{eq:em-int}\\
\int dp_{0}\frac{d\delta(p^{2})}{dp_{0}}p_{0}\partial_{i}f_{s} & = & -\int dp_{0}\delta(p^{2})\partial_{i}f_{s}-\int dp_{0}\delta(p^{2})p_{0}\frac{d}{dp_{0}}\partial_{i}f_{s}\nonumber \\
 & = & -\int dp_{0}\delta(p^{2})\partial_{i}f_{s}\nonumber \\
 &  & -\frac{1}{(2\pi)^{3}}\frac{d}{dE_{p}}\partial_{i}^{x}\left[f_{\mathrm{FD}}(E_{p}-\mu_{s})+f_{\mathrm{FD}}(E_{p}+\mu_{s})\right],\;(E_{p}^{0}\:\mathrm{in\: integrand})\nonumber \\
 & \rightarrow & -\frac{1}{(2\pi)^{3}}\frac{1}{E_{p}}\partial_{i}^{x}\left[f_{\mathrm{FD}}(E_{p}-\mu_{s})+f_{\mathrm{FD}}(E_{p}+\mu_{s})\right]\nonumber \\
 &  & +\frac{2}{(2\pi)^{3}}\frac{1}{E_{p}}\partial_{i}^{x}\left[f_{\mathrm{FD}}(E_{p}-\mu_{s})+f_{\mathrm{FD}}(E_{p}+\mu_{s})\right]\nonumber \\
 & = & \frac{1}{(2\pi)^{3}}\frac{1}{E_{p}}\partial_{i}^{x}\left[f_{\mathrm{FD}}(E_{p}-\mu_{s})+f_{\mathrm{FD}}(E_{p}+\mu_{s})\right],\label{eq:o-int-1}\\
\int dp_{0}\frac{d\delta(p^{2})}{dp_{0}}\frac{1}{p_{0}}\partial_{i}f_{s} & = & \int dp_{0}\delta(p^{2})\frac{1}{p_{0}^{2}}\partial_{i}f_{s}-\int dp_{0}\delta(p^{2})\frac{1}{p_{0}}\frac{d}{dp_{0}}\partial_{i}f_{s}\nonumber \\
 & = & \frac{1}{(2\pi)^{3}}\frac{1}{E_{p}^{3}}\partial_{i}^{x}\left[f_{\mathrm{FD}}(E_{p}-\mu_{s})+f_{\mathrm{FD}}(E_{p}+\mu_{s})\right]\nonumber \\
 &  & -\frac{1}{(2\pi)^{3}}\frac{1}{E_{p}^{2}}\frac{d}{dE_{p}}\partial_{i}^{x}\left[f_{\mathrm{FD}}(E_{p}-\mu_{s})+f_{\mathrm{FD}}(E_{p}+\mu_{s})\right],\;(E_{p}^{2}\:\mathrm{in\: integrand})\nonumber \\
 & \rightarrow & \frac{1}{(2\pi)^{3}}\frac{3}{E_{p}^{3}}\partial_{i}^{x}\left[f_{\mathrm{FD}}(E_{p}-\mu_{s})+f_{\mathrm{FD}}(E_{p}+\mu_{s})\right].\label{eq:o-int-2}
\end{eqnarray}
In Eqs. (\ref{eq:em-int-ix-b},\ref{eq:o-int-1},\ref{eq:o-int-2})
we have implied integrals over 3-momentum besides those over $p_{0}$,
so we can drop the complete derivative terms in $E_{p}$ which are
vanishing at two limits $E_{p}=0,\infty$.

\subsection{Evaluation of $I_{p0}$}

We now work on $I_{p0}=I_{p0}(\mathrm{EM})+I_{p0}(\omega)$, where
$I_{p0}(\mathrm{EM})$ and $I_{p0}(\omega)$ denote the first order
contribution from electromagnetic field and vorticity, respectively.
The results are 
\begin{eqnarray}
I_{p0}(\mathrm{EM}) & = & Q(\mathbf{p}\cdot\mathbf{E})\int dp_{0}\delta(p^{2})\partial_{0}^{p}f_{s}+\frac{1}{2}sQ^{2}(\mathbf{E}\cdot\mathbf{B})\int dp_{0}\frac{d\delta(p^{2})}{dp_{0}}\partial_{0}^{p}f_{s}\nonumber \\
 & = & \frac{1}{(2\pi)^{3}}Q(\mathbf{p}\cdot\mathbf{E})\frac{1}{E_{p}}\frac{d}{dE_{p}}\left[f_{\mathrm{FD}}(E_{p}-\mu_{s})-f_{\mathrm{FD}}(E_{p}+\mu_{s})\right]\nonumber \\
 &  & +\frac{1}{(2\pi)^{3}}sQ^{2}(\mathbf{E}\cdot\mathbf{B})\frac{1}{2E_{p}^{2}}\frac{d}{dE_{p}}\left[f_{\mathrm{FD}}(E_{p}-\mu_{s})+f_{\mathrm{FD}}(E_{p}+\mu_{s})\right],\label{eq:ip0-em}\\
I_{p0}(\omega) & = & -sQ\int dp_{0}\delta(p^{2})\left[\left(\frac{1}{2}-C_{10}-C_{11}\frac{p_{0}^{2}}{p^{2}}\right)(\mathbf{E}\cdot\boldsymbol{\omega})\right.\nonumber \\
 &  & \left.+(C_{30}+1)\frac{1}{p^{2}}(\mathbf{p}\cdot\boldsymbol{\omega})(\mathbf{p}\cdot\mathbf{E})\right]\partial_{0}^{p}f_{s}\nonumber \\
 & = & -sQ\left(\frac{1}{2}-C_{10}\right)(\mathbf{E}\cdot\boldsymbol{\omega})\int dp_{0}\delta(p^{2})\partial_{0}^{p}f_{s}\nonumber \\
 &  & -\frac{1}{2}sQC_{11}(\mathbf{E}\cdot\boldsymbol{\omega})\int dp_{0}\frac{d\delta(p^{2})}{dp_{0}}p_{0}\partial_{0}^{p}f_{s}\nonumber \\
 &  & +\frac{1}{2}sQ(C_{30}+1)(\mathbf{p}\cdot\boldsymbol{\omega})(\mathbf{p}\cdot\mathbf{E})\int dp_{0}\frac{d\delta(p^{2})}{dp_{0}}\frac{1}{p_{0}}\partial_{0}^{p}f_{s}\nonumber \\
 & = & \frac{1}{(2\pi)^{3}}sQ(C_{30}+1)\left[-\frac{1}{2E_{p}}(\mathbf{E}\cdot\boldsymbol{\omega})+\frac{3}{2E_{p}^{3}}(\mathbf{p}\cdot\boldsymbol{\omega})(\mathbf{p}\cdot\mathbf{E})\right]\nonumber \\
 &  & \frac{d}{dE_{p}}\left[f_{\mathrm{FD}}(E_{p}-\mu_{s})-f_{\mathrm{FD}}(E_{p}+\mu_{s})\right],\label{eq:ip0}
\end{eqnarray}
where we have used Eq. (\ref{eq:c0-c2}) in $I_{p0}(\omega)$. We
list following integrals involved in $I_{p0}(\mathrm{EM})$ and $I_{p0}(\omega)$,
\begin{eqnarray}
\int dp_{0}\frac{d\delta(p^{2})}{dp_{0}}\partial_{0}^{p}f_{s} & = & -\int dp_{0}\delta(p^{2})\frac{d}{dp_{0}}\partial_{0}^{p}f_{s}\nonumber \\
 & = & -\frac{1}{(2\pi)^{3}}\frac{1}{E_{p}}\frac{d^{2}}{dE_{p}^{2}}\left[f_{\mathrm{FD}}(E_{p}-\mu_{s})+f_{\mathrm{FD}}(E_{p}+\mu_{s})\right],\;(E_{p}^{0}\:\mathrm{in\: integrand})\nonumber \\
 & = & \frac{1}{(2\pi)^{3}}\frac{1}{E_{p}^{2}}\frac{d}{dE_{p}}\left[f_{\mathrm{FD}}(E_{p}-\mu_{s})+f_{\mathrm{FD}}(E_{p}+\mu_{s})\right],\\
\int dp_{0}\frac{d\delta(p^{2})}{dp_{0}}p_{0}\partial_{0}^{p}f_{s} & = & -\int dp_{0}\delta(p^{2})\partial_{0}^{p}f_{s}-\int dp_{0}p_{0}\delta(p^{2})\frac{d}{dp_{0}}\partial_{0}^{p}f_{s}\nonumber \\
 & = & -\frac{1}{(2\pi)^{3}}\frac{1}{E_{p}}\frac{d}{dE_{p}}\left[f_{\mathrm{FD}}(E_{p}-\mu_{s})-f_{\mathrm{FD}}(E_{p}+\mu_{s})\right]\nonumber \\
 &  & -\frac{1}{(2\pi)^{3}}\frac{d^{2}}{dE_{p}^{2}}\left[f_{\mathrm{FD}}(E_{p}-\mu_{s})-f_{\mathrm{FD}}(E_{p}+\mu_{s})\right],\;(E_{p}^{0}\:\mathrm{in\: integrand})\nonumber \\
 & \rightarrow & \frac{1}{(2\pi)^{3}}\frac{1}{E_{p}}\frac{d}{dE_{p}}\left[f_{\mathrm{FD}}(E_{p}-\mu_{s})-f_{\mathrm{FD}}(E_{p}+\mu_{s})\right]\nonumber \\
 & = & \int dp_{0}\delta(p^{2})\partial_{0}^{p}f_{s},\\
\int dp_{0}\frac{d\delta(p^{2})}{dp_{0}}\frac{1}{p_{0}}\partial_{0}^{p}f_{s} & = & \int dp_{0}\delta(p^{2})\frac{1}{p_{0}^{2}}\partial_{0}^{p}f_{s}-\int dp_{0}\delta(p^{2})\frac{1}{p_{0}}\frac{d}{dp_{0}}\partial_{0}^{p}f_{s}\nonumber \\
 & = & \frac{1}{(2\pi)^{3}}\frac{1}{E_{p}^{3}}\frac{d}{dE_{p}}\left[f_{\mathrm{FD}}(E_{p}-\mu_{s})-f_{\mathrm{FD}}(E_{p}+\mu_{s})\right]\nonumber \\
 &  & -\frac{1}{(2\pi)^{3}}\frac{1}{E_{p}^{2}}\frac{d^{2}}{dE_{p}^{2}}\left[f_{\mathrm{FD}}(E_{p}-\mu_{s})-f_{\mathrm{FD}}(E_{p}+\mu_{s})\right],\;(E_{p}^{2}\:\mathrm{in\: integrand})\nonumber \\
 & \rightarrow & \frac{3}{E_{p}^{3}}\frac{1}{(2\pi)^{3}}\frac{d}{dE_{p}}\left[f_{\mathrm{FD}}(E_{p}-\mu_{s})-f_{\mathrm{FD}}(E_{p}+\mu_{s})\right],
\end{eqnarray}
where we have implied that there will be integrals over $\mathbf{p}$
besides those over $p_{0}$ so we have dropped the complete derivatives
in the integral $\int_{0}^{\infty}dE_{p}$ (with $E_{p}=|\mathbf{p}|$)
since they are vanishing at two limits $E_{p}=0,\infty$.

\subsection{Evaluation of $I_{p}$}

Finally we work on $I_{p}=I_{p}(\mathrm{EM})+I_{p}(\omega)$, where
$I_{p}(\mathrm{EM})$ and $I_{p}(\omega)$ denote the zeroth order
contribution, the first order contribution from electromagnetic field
and vorticity, respectively. Now we evaluate $I_{p}(\mathrm{EM})$
and $I_{p}(\omega)$, 
\begin{eqnarray}
I_{p}(\mathrm{EM}) & = & Q\int dp_{0}\delta(p^{2})[p_{0}\mathbf{E}_{i}+(\mathbf{p}\times\mathbf{B})_{i}]\partial_{i}^{p}f_{s}\nonumber \\
 &  & +\frac{1}{2}sQ^{2}(\mathbf{E}\cdot\mathbf{B})\mathbf{p}_{i}\int dp_{0}\frac{d\delta(p^{2})}{dp_{0}}\frac{1}{p_{0}}\partial_{i}^{p}f_{s}\nonumber \\
 & = & \frac{1}{(2\pi)^{3}}Q\mathbf{E}_{i}\partial_{i}^{p}\left[f_{\mathrm{FD}}(E_{p}-\mu_{s})-f_{\mathrm{FD}}(E_{p}+\mu_{s})\right]\nonumber \\
 &  & +\frac{1}{(2\pi)^{3}}\left[Q\mathbf{E}+Q\frac{\mathbf{p}}{E_{p}}\times\mathbf{B}+sQ^{2}(\mathbf{E}\cdot\mathbf{B})\frac{\mathbf{p}}{E_{p}^{3}}\right]_{i}\nonumber \\
 &  & \times\partial_{i}^{p}\left[f_{\mathrm{FD}}(E_{p}-\mu_{s})+f_{\mathrm{FD}}(E_{p}+\mu_{s})\right],\label{eq:ip-em}\\
I_{p}(\omega) & = & -sQ\int dp_{0}\delta(p^{2})\frac{1}{p^{2}}p_{0}(\mathbf{p}\cdot\boldsymbol{\omega})\mathbf{E}_{i}\partial_{i}^{p}f_{s}\nonumber \\
 &  & +sQ\int dp_{0}\delta(p^{2})\mathbf{p}_{i}\left[-C_{40}(\boldsymbol{\omega}\cdot\mathbf{E})\frac{1}{p_{0}}+C_{41}\frac{1}{p^{2}p_{0}}(\mathbf{p}\cdot\boldsymbol{\omega})(\mathbf{p}\cdot\mathbf{E})\right]\partial_{i}^{p}f_{s}\nonumber \\
 & = & \frac{1}{2}sQ(\mathbf{p}\cdot\boldsymbol{\omega})\mathbf{E}_{i}\int dp_{0}\frac{d\delta(p^{2})}{dp_{0}}\partial_{i}^{p}f_{s}\nonumber \\
 &  & -sQC_{40}(\boldsymbol{\omega}\cdot\mathbf{E})\mathbf{p}_{i}\int dp_{0}\delta(p^{2})\frac{1}{p_{0}}\partial_{i}^{p}f_{s}\nonumber \\
 &  & -\frac{1}{2}sQC_{41}(\mathbf{p}\cdot\boldsymbol{\omega})(\mathbf{p}\cdot\mathbf{E})\mathbf{p}_{i}\int dp_{0}\frac{d\delta(p^{2})}{dp_{0}}\frac{1}{p_{0}^{2}}\partial_{i}^{p}f_{s}\nonumber \\
 & = & \frac{1}{(2\pi)^{3}}sQ\frac{1}{E_{p}^{2}}(\mathbf{p}\cdot\boldsymbol{\omega})\mathbf{E}_{i}\partial_{i}^{p}\left[f_{\mathrm{FD}}(E_{p}-\mu_{s})-f_{\mathrm{FD}}(E_{p}+\mu_{s})\right]\nonumber \\
 &  & -\frac{1}{(2\pi)^{3}}C_{40}sQ\frac{1}{E_{p}^{2}}(\boldsymbol{\omega}\cdot\mathbf{E})\mathbf{p}_{i}\partial_{i}^{p}\left[f_{\mathrm{FD}}(E_{p}-\mu_{s})-f_{\mathrm{FD}}(E_{p}+\mu_{s})\right]\nonumber \\
 &  & -\frac{1}{(2\pi)^{3}}C_{41}sQ\frac{2}{E_{p}^{4}}(\mathbf{p}\cdot\boldsymbol{\omega})(\mathbf{p}\cdot\mathbf{E})\mathbf{p}_{i}\partial_{i}^{p}\left[f_{\mathrm{FD}}(E_{p}-\mu_{s})-f_{\mathrm{FD}}(E_{p}+\mu_{s})\right]\nonumber \\
 & = & \frac{1}{(2\pi)^{3}}sQ\left[\frac{1}{E_{p}^{2}}(\mathbf{p}\cdot\boldsymbol{\omega})\mathbf{E}_{i}-C_{40}\frac{1}{E_{p}^{2}}(\boldsymbol{\omega}\cdot\mathbf{E})\mathbf{p}_{i}-C_{41}\frac{2}{E_{p}^{4}}(\mathbf{p}\cdot\boldsymbol{\omega})(\mathbf{p}\cdot\mathbf{E})\mathbf{p}_{i}\right]\nonumber \\
 &  & \times\partial_{i}^{p}\left[f_{\mathrm{FD}}(E_{p}-\mu_{s})-f_{\mathrm{FD}}(E_{p}+\mu_{s})\right],\label{eq:ip-vort}
\end{eqnarray}
where we have used 
\begin{eqnarray}
\int dp_{0}\frac{d\delta(p^{2})}{dp_{0}}\frac{1}{p_{0}}\partial_{i}^{p}f_{s} & = & \int dp_{0}\delta(p^{2})\frac{1}{p_{0}^{2}}\partial_{i}^{p}f_{s}-\int dp_{0}\delta(p^{2})\frac{1}{p_{0}}\frac{d}{dp_{0}}\partial_{i}^{p}f_{s}\nonumber \\
 & = & \frac{1}{(2\pi)^{3}}\frac{1}{E_{p}^{3}}\partial_{i}^{p}\left[f_{\mathrm{FD}}(E_{p}-\mu_{s})+f_{\mathrm{FD}}(E_{p}+\mu_{s})\right]\nonumber \\
 &  & -\frac{1}{(2\pi)^{3}}\frac{1}{E_{p}^{2}}\frac{d}{dE_{p}}\partial_{i}^{p}\left[f_{\mathrm{FD}}(E_{p}-\mu_{s})+f_{\mathrm{FD}}(E_{p}+\mu_{s})\right],\;(E_{p}\:\mathrm{in\: integrand})\nonumber \\
 & = & \frac{1}{(2\pi)^{3}}\frac{2}{E_{p}^{3}}\partial_{i}^{p}\left[f_{\mathrm{FD}}(E_{p}-\mu_{s})+f_{\mathrm{FD}}(E_{p}+\mu_{s})\right],\\
\int dp_{0}\frac{d\delta(p^{2})}{dp_{0}}\partial_{i}^{p}f_{s} & = & -\int dp_{0}\delta(p^{2})\frac{d}{dp_{0}}\partial_{i}^{p}f_{s}\nonumber \\
 & = & -\frac{1}{(2\pi)^{3}}\frac{1}{E_{p}}\frac{d}{dE_{p}}\partial_{i}^{p}\left[f_{\mathrm{FD}}(E_{p}-\mu_{s})-f_{\mathrm{FD}}(E_{p}+\mu_{s})\right],\;(E_{p}\:\mathrm{in\: integrand})\nonumber \\
 & = & \frac{1}{(2\pi)^{3}}\frac{2}{E_{p}^{2}}\partial_{i}^{p}\left[f_{\mathrm{FD}}(E_{p}-\mu_{s})-f_{\mathrm{FD}}(E_{p}+\mu_{s})\right],\\
\int dp_{0}\delta(p^{2})\frac{1}{p_{0}}\partial_{i}^{p}f_{s} & = & \frac{1}{(2\pi)^{3}}\frac{1}{E_{p}^{2}}\partial_{i}^{p}\left[f_{\mathrm{FD}}(E_{p}-\mu_{s})-f_{\mathrm{FD}}(E_{p}+\mu_{s})\right],\\
\int dp_{0}\frac{d\delta(p^{2})}{dp_{0}}\frac{1}{p_{0}^{2}}\partial_{i}^{p}f_{s} & = & 2\int dp_{0}\delta(p^{2})\frac{1}{p_{0}^{3}}\partial_{i}^{p}f_{s}-\int dp_{0}\delta(p^{2})\frac{1}{p_{0}^{2}}\frac{d}{dp_{0}}\partial_{i}^{p}f_{s}\nonumber \\
 & = & \frac{1}{(2\pi)^{3}}\frac{2}{E_{p}^{4}}\partial_{i}^{p}\left[f_{\mathrm{FD}}(E_{p}-\mu_{s})-f_{\mathrm{FD}}(E_{p}+\mu_{s})\right]\nonumber \\
 &  & -\frac{1}{(2\pi)^{3}}\frac{1}{E_{p}^{3}}\frac{d}{dE_{p}}\partial_{i}^{p}\left[f_{\mathrm{FD}}(E_{p}-\mu_{s})-f_{\mathrm{FD}}(E_{p}+\mu_{s})\right],\;(E_{p}^{3}\:\mathrm{in\: integrand})\nonumber \\
 & \rightarrow & \frac{1}{(2\pi)^{3}}\frac{4}{E_{p}^{4}}\partial_{i}^{p}\left[f_{\mathrm{FD}}(E_{p}-\mu_{s})-f_{\mathrm{FD}}(E_{p}+\mu_{s})\right].
\end{eqnarray}
Here we have implied that there will be integrals over $\mathbf{p}$
besides those over $p_{0}$ so we have dropped the complete derivatives
in the integral $\int_{0}^{\infty}dE_{p}$ (with $E_{p}=|\mathbf{p}|$)
since they are vanishing at two limits $E_{p}=0,\infty$. 

\bibliographystyle{apsrev}
\bibliography{ref}

\end{document}